\documentclass[aps,prd,preprint,showpacs,floatfix,preprintnumbers,nofootinbib,superscriptaddress,
showkeys]{revtex4}
\usepackage[utf8]{inputenc}
\usepackage{fancybox}
\usepackage{hhline}
\usepackage{dcolumn}
\usepackage{textcomp}
\usepackage{epsfig,graphics,graphicx}
\usepackage{amsfonts,amssymb,amsmath}
\usepackage{pifont}
\usepackage{bm}
\usepackage{longtable} 
\usepackage{appendix}
\usepackage{lscape}
\usepackage[mathscr]{euscript}
\usepackage{mathrsfs}
\usepackage{multirow}
\usepackage{rotating}
\usepackage{color}   %May be necessary if you want to color links

\newcommand{\beq}{\begin{equation}}
\newcommand{\eeq}{\end{equation}}
\newcommand{\bea}{\begin{eqnarray}}
\newcommand{\eea}{\end{eqnarray}}

\newcommand{\eps}{\epsilon}
\newcommand{\ord}[1]{{\cal{O}}( #1 )}
\newcommand{\B}{{\bf B}}
\newcommand{\Bdag}{{\bf B^\dagger}}

\newcommand{\gA}{\mathring g_A}

\DeclareFontFamily{OT1}{pzc}{}
\DeclareFontShape{OT1}{pzc}{m}{it} 
              {<-> s * [0.900] pzcmi7t}{}
\DeclareMathAlphabet{\mathpzc}{OT1}{pzc} 
                                 {m}{it}
\DeclareMathAlphabet{\mathcalligra}{T1}{calligra}{m}{n}
\usepackage{hyperref}
\hypersetup{pdfauthor={me}, colorlinks=true, citecolor=blue, urlcolor=blue, linkcolor=black}
\begin{document}
\preprint{\vbox{\hbox{ JLAB-THY-19-3088} }}
 
\title{  
 The SU(3) Vector Currents in BChPT $\!\boldsymbol{ \times \, {\rm  
  1/N  _c}}$   }
\author{I.~P.~Fernando}\email{ishara@jlab.org }
\author{J.~L.~Goity}\email{goity@jlab.org}
\affiliation{Department of Physics, Hampton University, Hampton, VA 23668, USA. }
\affiliation{Thomas Jefferson National Accelerator Facility, Newport News, VA 23606, USA.}
 
\begin{abstract}
Baryon Chiral Perturbation Theory  (BChPT)  combined with the  $1/N_c$ expansion  is applied to the $SU(3)$ vector currents.  In terms of the $\xi$ power counting   linking the low energy and $1/N_c$ expansions according to $\ord{\xi}=\ord{p}=\ord{1/N_c}$, the study  is carried out to next-to-next-to-leading order, and it
  includes   $SU(3)$ breaking corrections to the $\mid \! \Delta S\!\mid=1$ vector charges,     charge radii,  and magnetic moments and  radii.  The results are   obtained for generic $N_c$, allowing for investigating the various scalings in $N_c$. 
\end{abstract}
\maketitle

\section{Introduction}
\label{sec:Intro}

Vector currents, being intimately related to the  flavor $SU(3)$  symmetry of QCD, represent a fundamental probe for hadron structure as well as for the breaking of     $SU(3)$ by quark masses. This is particularly interesting for baryons, where the electromagnetic current for   nucleons, known empirically to remarkable accuracy \cite{Ye:2017gyb}, along  with the magnetic moments of hyperons allow for an almost complete   description of all the $SU(3)$ vector currents to the order  in the low energy expansion considered in the  present work.
The charged vector currents are relevant in $\beta$ decays, where both   $SU(3)$ breaking in the $\mid \!  \Delta S \! \mid=1$ charges and weak magnetism are still open problems. To the present level of experimental accuracy in hyperon $\beta$ decays, there is not sufficient  sensitivity to  the  $SU(3)$ breaking in the charges  \cite{Cabibbo:2003cu}. The reason  is that $\beta$ decay has a branching fraction of about $10^{-3}$,  being dominated by the non-leptonic component. Fortunately,  lattice QCD is producing  results \cite{Aubin:2008qp,
Shanahan:2015dka,Parreno:2016fwu,Alexandrou:2018zdf} which can be compared with the predictions of the  approach in the present work.    The experimental information on charge form factors is  limited to the electric  form factors of nucleons and  the charge radius of the $\Sigma^-$. This is however sufficient to predict the rest of the charge radii, whose $SU(3)$ breaking effects are, at the order of the present calculation,    finite    non-analytic  in quark masses. The  octet baryons'  EM magnetic moments and nucleons' magnetic radii  give an almost complete prediction for the rest of the currents but for one low energy constant (LEC) which requires knowledge of at least one weak magnetic moment of a $\Delta S=1$ current. 
 In the approach   followed here, results automatically extend to the vector current observables of the decuplet baryons and to EM transitions, e.g.  the $M_1$ transition $\Delta\to N\gamma$,   most of which  remain  empirically unknown or poorly known. 
 The study of electric currents in BChPT with inclusion of the spin 3/2 baryons dates   back   a quarter century \cite{Jenkins:1992pi,Butler_1994}, and numerous works have since been produced in various versions of that framework, among those close in spirit to  the present  one are found in Refs. \cite{Durand:1997ya,Bernard:1998gv,Pascalutsa:2004je,Ledwig:2011cx,Geng:2009ys,Geng:2009hh,Jiang:2009jn}, and works with additional  constraints imposed by consistency with the $1/N_c$  expansion are those of Refs. \cite{Luty:1994ub,FloresMendieta:2009rq,Ahuatzin:2010ef,Jenkins:2011dr,Flores-Mendieta:2014vaa,Flores-Mendieta:2015wir}. The present work formalizes the combination of BChPT and the $1/N_c$ expansion  \cite{Jenkins:1995gc} for the vector currents following the rigorous power counting scheme of the $\xi$ expansion \cite{CalleCordon:2012xz,Fernando:2017yqd}  based on the linking $\ord{p}=\ord{1/N_c}=\ord{\xi}$. 
The combined framework  was first applied to the $SU(3)$ vector charges in Ref. \cite{Flores-Mendieta:2014vaa}, where  the  $\xi$ expansion     was not strictly implemented, however for the purpose  of calculating  the corrections of $SU(3)$ breaking to the vector charges,  restricted by the  Ademollo-Gatto theorem (AGT), such omission has no very significant effect \footnote{In \cite{Flores-Mendieta:2014vaa} the baryon-GB vertices included higher order terms in $1/N_c$.}. 
Here a  complete   study  is presented to $\ord{\xi^3}$ and $\ord{\xi^4}$ (depending on the observable) of the $SU(3)$ vector currents. The present work provides results for generic $N_c$, permitting in this way to sort out in particular the large $N_c$ behavior of non-analytic terms in $\xi$  stemming from   one loop corrections,  which gives additional understanding, as it has been   shown for instance in the case of the Gell-Mann-Okubo relation and  the $\sigma$ terms discussed in Refs. \cite{Fernando:2018jrz,Fernando:2019sfm}.
The subject of magnetic moments has been addressed in the context of the $1/N_c$ expansion in   works limited to a tree level   expansion in composite operators \cite{Jenkins_1994,Dai:1995zg,Buchmann:2002et,Lebed:2004fj}, and in works including one loop corrections in BChPT   Refs. \cite{FloresMendieta:2009rq,Ahuatzin:2010ef,Jenkins:2011dr,Flores-Mendieta:2015wir}. 
 In addition to the BChPT,  dispersive approaches implementing constraints of chiral dynamics have been implemented \cite{Hammer:2003ai,Belushkin:2005ds,Belushkin:2006qa} and where in addition consistency with the $1/N_c$ expansion has also been required \cite{Granados:2013moa,Granados:2016jjl,Alarcon:2017asr,Alarcon:2017lhg,Alarcon:2018irp}. Such works  naturally give a   range of applicability beyond the present one, which is limited up to the form factor  radii. 
 
 This work is organized as follows: Section II presents the baryon chiral Lagrangians needed for the present work, Section III   summarizes the one loop corrections to the vector currents, Section IV presents the analysis of the vector charges and radii, and Section V does the same for the magnetic moments and radii. A summary is presented in Section VI. Several appendices are included for the benefit of readers intending to implement similar calculations.
 
\section{Baryon Chiral Lagrangian}
 
This section summarizes the pieces of the Baryon Chiral Lagrangian up to $\ord{\xi^4}$ relevant to the calculations in this work.  The details on the construction of the Lagrangians and the notations are given in Ref. \cite{Fernando:2017yqd}. 
In order to ensure the validity of the OZI rule for the quark mass dependency of baryon masses, namely, that the non-strange baryon mass dependence on $m_s$ be  $\ord{N_c^0}$, the following combination of the source $\chi_+$ is defined by \cite{Fernando:2017yqd}:
\beq
\hat\chi_+\equiv \tilde\chi_+ +N_c\; \chi_+^0,
\eeq
which is $\ord{N_c}$ but has dependence on $m_s$ which is $\ord{N_c^0}$  for all states with strangeness   $\ord{N_c^0}$. For convenience a scale $\Lambda$ is introduced, which  can be chosen to be a typical  QCD scale, in order to render most of the LECs dimensionless. In the calculations $\Lambda=m_\rho$ will be chosen. The quark mass matrix is defined by ${\cal M}_q=m^0+m^a \frac{\lambda^a}{2}$, where in the physical case $m^0=\frac 13 (m_u+m_d+m_s)$, $m^3=m_u-m_d$ and $m^8=\frac{1}{\sqrt{3}} (m_u+m_d-2 m_s)$ and the rest of the $M^a$ s vanish.

Collecting the baryons in a   spin flavor multiplet denoted by $\B$, and using standard notation for the chiral building blocks (for details see \cite{Fernando:2017yqd}), the LO $\ord{\xi}$ Lagrangian reads:
\beq
{\cal{L}}_B^{(1)}=\Bdag (i D_0-\frac{C_{H\!F}}{N_c} \hat S^2+\gA  u^{ia} G^{ia}+\frac{c_1}{2\Lambda}\hat \chi_+)\B,
\label{eqn:LOLagrangian}
\eeq
where the hyperfine mass shifts are given by the second term, $G^{ia}$ are the spin-flavor generators (see Appendix \ref{app:Algebra}), and the axial coupling is at LO $\gA=\frac 65 g_A$, being $g_A=1.2732(23)$  the nucleon's axial coupling.
The relevant terms in  the $\ord{\xi^2}$ Lagrangian are:
\bea
{\cal{L}}_B^{(2)}&=&\Bdag \left(   \frac{c_2}{\Lambda}\chi^0_++ \frac{C_1^A}{N_c} u^{ia} S^i T^a +\frac{\kappa}{2 \Lambda}  B_+^{ia} G^{ia}+\cdots\right)\B,
\eea
where the flavor $SU(3)$ electric and magnetic fields are denoted by $E_+$ and $B_+$   and   given by $E_+^i=F_+^{0i}$ and $B_+^i=\frac 12 \eps^{ijk} F_+^{jk}$ \cite{Fernando:2017yqd}. The term proportional to  $\kappa$ gives  at LO the   magnetic moments associated with all vector currents. 
 The 
$\ord{\xi^3}$ and $\ord{\xi^4}$ Lagrangians   needed for the one-loop renormalization of the vector currents  are the following:
\bea
{\cal{L}}_B^{(3)}&=&\Bdag \Big(\frac{g_1}{\Lambda^2}D_i E_{+i}^a T^a+ \frac{\kappa_1}{2\Lambda N_c} B_+^{ia} S^i T^a+\cdots \Big)\B \nonumber\\
{\cal{L}}_B^{(4)}&=&\Bdag  \left(\frac{1}{N_c\Lambda^2}(g_2 D_i E_{+i}^a S^jG^{ja}+g_3  D_i E_{+j}^a \{S^i,G^{ja}\}^{\ell=2})+\frac{\kappa_r}{\Lambda^3} D^2 B_+^{ia} G^{ia}\right.\nonumber\\
&+&\left.\frac{1}{2\Lambda^3} (\kappa_2 \chi_+^0 B_+^{ia} G^{ia}+i \kappa_F f^{abc} \chi_+^a B_+^{ib} G^{ic}+\kappa_D d^{abc} \chi_+^a B_+^{ib} G^{ic}+\kappa_3 \chi_+^a B_+^{ia} S^i)\right.\nonumber\\
&+&\left. \frac{1}{2 \Lambda N_c^2} (\kappa_4 B_+^{ia}\{\hat S^2,G^{ia}\}+\kappa_5   B_+^{ia}S^i S^j G^{ja})+\cdots \right)\B   
\eea
The LECs  $g_1$ and $g_2$ will be determined by charge radii, the term proportional to $g_3$ gives electric quadrupole moments for decuplet baryons and  for  transitions between decuplet to octet baryons, which will not be discussed here, and the term proportional to $\kappa_r$ gives a contribution to  magnetic radii ($D^2 B_+\equiv D_\mu D^\mu B_+ $ being the covariant divergence of the magnetic field). The rest are quark mass and  higher order   in $1/N_c$  corrections to the magnetic moments.

    Throughout,  spin-flavor operators   in the   Lagrangians are scaled by  appropriate powers of $1/N_c$ such   that all   LECs start at zeroth order in $N_c$. Of course, LECs have themselves an expansion in $1/N_c$,  kept implicit,  which requires  information  for  $N_c>3$ to be determined. In that sense  each Lagrangian term has a leading power in $1/N_c$ which is used to assign its order in the $\xi$ power counting, followed by sub-leading terms in $1/N_c$ due to the expansion of the corresponding LEC.  In addition,  each term in the Lagrangian is explicitly chiral invariant  and its expansion in powers of the Goldstone Boson fields yields factors $1/F_\pi=\ord{1/\sqrt{N_c}}$ for each additional factor of a GB field. 
    
 For convenience the following definition is used:
 \bea
 \delta \hat m&\equiv&  \frac{C_{\rm HF}}{N_c}{ \hat{S}^2}-\frac{c_1}{2\Lambda}\; \hat\chi_+.
    \label{eq:deltam}
 \eea
  Note that $\delta \hat m$ gives rise to mass splittings between baryons which are the $\ord{1/N_c}$ hyperfine term in  Eqn.(\ref {eqn:LOLagrangian}) and the $\ord{p^2}$ quark mass term. The $\ord{m_q N_c}$ term in    $\hat\chi_+$   becomes immaterial in the loop calculations as only differences of baryon masses appear for which such terms exactly cancel.

 \section{One loop corrections to currents}
 The one-loop corrections to  the vector currents  involve the two sets of gauge invariant diagrams $A$ and $B$   in Fig. \ref{fig:1-loop-VC}, where the vertices   are  given  in Appendix \ref{app:vertices}.   The explicit results are the following: 
\bea
V^{\mu a}( A_1) &=&i  \left( \frac{\mathring g_A}{F_\pi} \right)^2 \sum_{n_1,n_2} G^{ib} {\cal{P}}_{n_2} \Gamma^{\mu a}  {\cal{P}}_{n_1} G^{jb}\;\frac{1}{q_0-\delta m_{n_2}+\delta m_{n_1}}\nonumber\\
&\times&\left(H^{ij}(p_0-\delta m_{n_1},M_b)-H^{ij}(p_0+q_0-\delta m_{n_2},M_b)\right)\nonumber\\
V^{\mu a}(A_2) &=& \frac 12 \{ \Gamma^{\mu a},\delta \hat Z_{1-loop}\}\nonumber\\
V^{\mu a}(A_3) &=& \left( \frac{\mathring g_A}{F_\pi} \right)^2 f^{abc} \sum_{n} G^{ib}   {\cal{P}}_{n} G^{jc} H^{ij\mu} (p_0-\delta m_n, q, M_b,M_c)\nonumber\\
V^{\mu a}(B_1) &=& - \frac{i}{2F_\pi^2} f^{abc}f^{bcd} \Gamma^{\mu d} I(0,1,M_b^2) \\
V^{\mu a}(B_2 )&=&g^{\mu 0}\frac{i}{4F_\pi^2} f^{abc}f^{bcd} T^d({q_0}^2 K(q,M_b,M_c)+4 q_0 K^0(q,M_b,M_c)+4 K^{00}(q,M_b,M_c)) ,\nonumber
\label{eqn:loopdiags}
\eea
where ${\cal{P}}_{n}$ are projectors onto the corresponding baryon in the loop, $p_0$ is the residual energy of the initial baryon,  $q_0$ is the incoming energy in the current, and $\Gamma^{\mu a}=g^{\mu 0} T^a+i\frac{\kappa}{\Lambda} \eps^{0\mu i j} f^{abc}f^{cbd}q^i G^{jd}$ contains both the electric charge and magnetic moment components.
 The one-loop  wave function renormalization factor $\delta \hat Z_{1-loop}$ can be found in \cite{Fernando:2017yqd}, and the loop integrals  $I$, $K$, $K^\mu$, $K^{\mu\nu}$, $H^{ij}$ and $H^{ij\mu}$ are given in Appendix \ref{app:loopintegrals}.  Since the temporal component of the current can only connect baryons with the same spin,   $q_0$ is equal to the $SU(3)$ breaking mass difference between them plus the kinetic energy transferred by the current, which are all $\ord{\xi^2}$ or higher and  must  therefore  be neglected in this calculation. In the evaluations one sets  $p_0\to \delta m_{in}$ and $p_0+q_0\to \delta m_{out}$. In particular,  for  diagram $A_1$, if it requires evaluation at $q_0=0$ such a limit must be taken in the end of the evaluation.   The $U(1)$ baryon number current can used to check the calculation: only diagrams $A_{1+2}$ contribute, and as required they cancel each other. 
\begin{center}
\begin{figure}[h]
\includegraphics[height=5cm,width=12cm,angle=0]{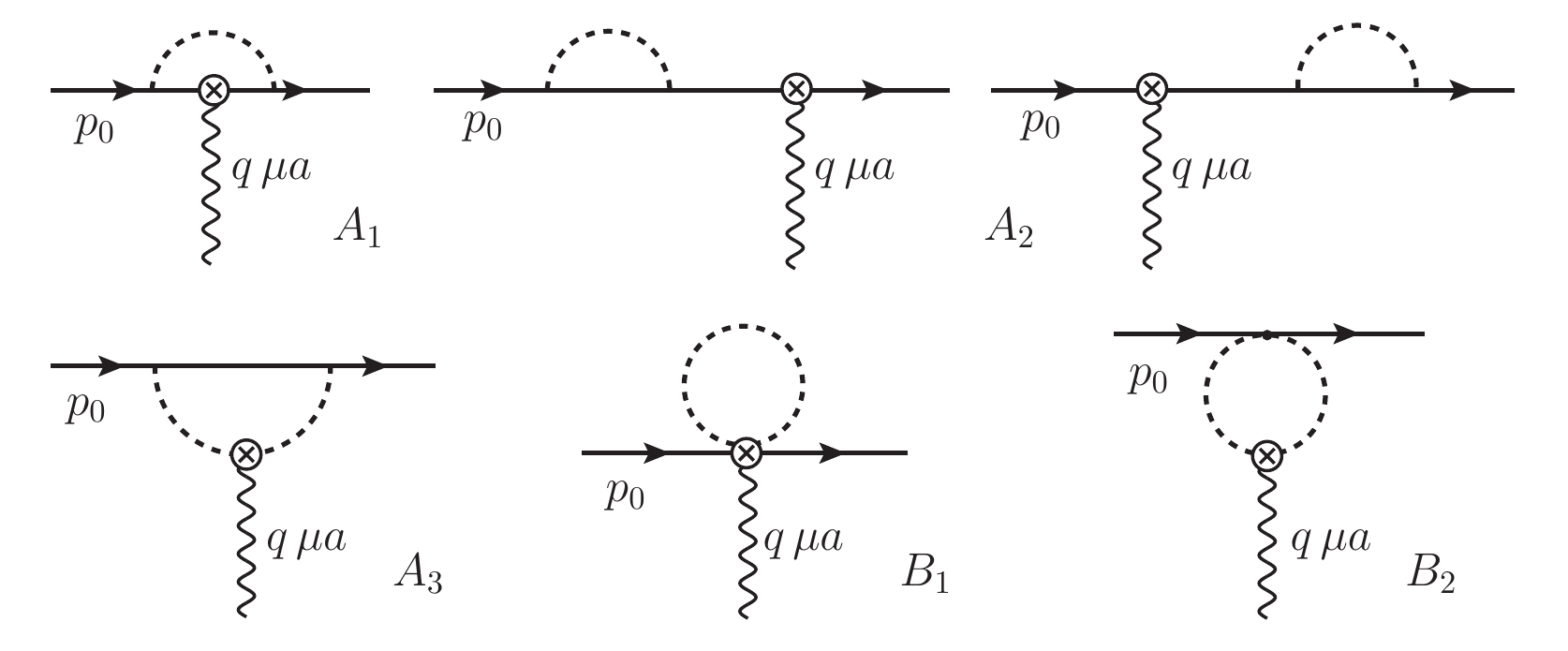}  
\caption{Diagrams contributing to the 1-loop corrections to the vector currents. }
\label{fig:1-loop-VC}
\end{figure}
\end{center}

For a generic current vertex $\Gamma$, the combined UV divergent and polynomial piece of diagrams  $A_{1+2}$ can be written as:
\bea
\Gamma(A_{1+2})^{\rm poly}&=&\frac{1}{(4\pi)^2}    \left( \frac{\mathring g_A}{F_\pi} \right)^2 \left( \frac 12 (\lambda_\eps+1) M_{ab}^2[G^{ia},[G^{ib},\Gamma]]\right.\nonumber\\
&+&\left. \frac 13(\lambda_\eps+2) \left(2[[G^{ia},\Gamma],[\delta\hat m,[\delta\hat m,G^{ia}]]]+[[\Gamma,[\delta\hat m,G^{ia}]],[\delta\hat m,G^{ia}]]\right)\right),
\label{eqn:A12UV}
\eea 
where $\lambda_\eps=\frac 1\eps-\gamma+\log  4\pi$.
The first term is proportional to quark masses through the GB mass-square matrix $M_{ab}^2=m^0 \delta^{ab}+\frac 12 d^{abc} m^c $, and the second involves the baryon  hyperfine mass splittings $\delta\hat m$ which are $\ord{1/N_c}$ and, following the strict $\xi$ power counting,   the $\ord{p^2}$ terms due to $SU(3)$ breaking in $\delta\hat m$  are disregarded. The consistency with the $1/N_c$ power counting  can be readily checked. Diagrams $A_3$ and $B_{1,2}$ are separately  consistent with the $1/N_c$ power counting. Their polynomial contributions  are the following: \bea
V^{\mu a}(A_3)^{Poly}&=&-  \frac{1}{(4\pi)^2}  \left( \frac{\mathring g_A}{F_\pi} \right)^2  \frac 16 \;i f^{abc} 
\nonumber\\ &\times& 
\left(g^{\mu 0} \Big ((\lambda_\eps q^i q^j+\frac 12 (\lambda_\eps+1) q^2 g^{ij})\delta^{bd}-3g^{ij}(\lambda_\eps +1) M_{bd}^2\Big)[G^{id}, G^{jc}] \right. \nonumber\\
&-&g^{\mu 0}\left.(\lambda_\eps+2) \Big(\frac 12 [G^{ib} ,[[G^{ic},\delta \hat m],\delta\hat m]]
-[[G^{ib},\delta\hat m],[G^{ic},\delta\hat m]]\Big)\right.
\nonumber\\
&+&\left. g^\mu_i  (\lambda_\eps+2)\Big( \frac 12 g^{jk} q^i [[G^{kb} ,G^{jc}],\delta\hat m]+2 g^{ij} q^k (3[G^{kb},[G^{jc},\delta \hat m]]+[[G^{jc},G^{kb}],\delta\hat m])\Big)\right)
\nonumber\\
V^{\mu a}(B_1)^{Poly}&=& \frac{1}{(4\pi)^2} (\lambda_\eps+1)  \frac{1}{2 F_\pi^2}   f^{abd}f^{cde} \,\Gamma^{\mu e} M_{bc}^2
\nonumber\\
&=& -  \frac{1}{(4\pi)^2} \frac{3}{F_\pi^2} (\lambda_\eps+1) g^{\mu 0} B_0(m^0 \,\Gamma^{\mu a}+\frac 1 4 d^{abc} m^b \,\Gamma^{\mu c})\nonumber\\
V^{\mu a}(B_2)^{Poly}&=&- \frac{1}{(4\pi)^2} \lambda_\eps \frac{1}{4 F_\pi^2} (g^{\mu 0} \,\vec q^{\;2} +g^\mu_iq^i q_0)\,T^a-g^{\mu 0}V^{0 a}(B_1)^{Poly} 
\label{eqn:A3B12UV}
\eea
Reduction formulas that can be found in \cite{Fernando:2018jrz} are used to express the above in a base of irreducible operators, Eqns.(\ref{eqn:chargeFFUV}) and (\ref{eqn:UVMagMom}) below.
 
 \section{Vector Charges}
\label{sec:charges}
In this section   the SU(3) vector current charges  and corresponding radii are analyzed. The SU(3) breaking corrections to the charges already presented in \cite{Flores-Mendieta:2014vaa} and \cite{Fernando:2017yqd} are discussed for completeness.   At lowest order the charges  are represented by  the flavor generators $T^a$. The   one-loop corrections  are UV finite at $Q^2\equiv -q^2=0$, and since up to $\ord{\xi^3}$ the  AGT is satisfied,  the corrections to the charges are unambiguously given by   UV finite one-loop contributions. 
Note that the AGT applies to the whole baryon spin-flavor multiplet.
On the other hand, at finite $Q^2$ the one-loop correction has an UV divergent piece which is independent of quark masses and is renormalized via the terms $g_1$ and $g_2$ in ${\cal{L}}_B $, one of them removes the UV divergence ($g_1$) and the other one is a finite counterterm ($g_2$).

Combining the polynomial pieces in Eqns.(\ref{eqn:A12UV}) and  (\ref{eqn:A3B12UV})  and using that, $[\delta \hat m,T^a]=[\delta \hat m,\hat G^2]=[\delta \hat m,G^{ib} T^a G^{ib}]=0$ one obtains the polynomial loop contributions to vector charges, which are proportional to $Q^2=\vec {q}^{\,2}$:
\bea
f_1^a(A_{1+2+3})^{\text{poly}}&=&\frac{\lambda_\eps-3}{(4\pi)^2}\left( \frac{\mathring g_A}{4F_\pi} \right)^2Q^{2} \,T^a\nonumber\\
f_1^a(B_{1+2})^{\text{poly}}&=&-\frac{\lambda_\eps+1}{(4\pi)^2} \frac{Q^2}{4 F_\pi^2} \,  T^a,
\label{eqn:chargeFFUV}
\eea
where $f_1^a\equiv V^{0a}$.
 
The   corrections  to the $\mid \Delta S\mid=1$ charges, already discussed in  \cite{Flores-Mendieta:2014vaa}, are evaluated using the physical values   $\gA=\frac 6 5 \times 1.27$ and $F_\pi=92$ MeV,   however  one needs to be aware that their values are effected by the NLO corrections, leading to a theoretical uncertainty. With the usual notation for those charges \cite{Flores-Mendieta:2014vaa}, evaluating the ratios $\delta f_1/f_1$  in the large  $N_c$ limit one finds that $\delta f_1/f_1=\ord{1/N_c}$. However, this behavior   sets in rather slowly at  $N_c\sim 20$, emphasizing the fact that the non commutativity of the low energy and $1/N_c$ expansions is very important   at the physical $N_c=3$. The results are shown in Table \ref{table:df1}, where the errors  are   estimated  from  the above mentioned theoretical uncertainty. The agreement with recent LQCD calculations \cite{Shanahan:2015dka} is encouraging, and further improvement in the precision of those  calculations would be very useful.

\begin{table}[htp]
%\label{table:df1}
\begin{center}
\begin{tabular}{lll}
\hline
&\multicolumn{2}{c}{$\frac{\delta f_1}{f_1} $}  \\
  ~~~~~& \multicolumn{1}{l}{ ~One-loop} & ~ LQCD \\
\hline
$\Lambda p$ &        $-0.067(15)$~~~~~   & $-0.05 (2)$        \\
$\Sigma^- n$ &    $-0.025(10)$     &$-0.02 (3)$      \\
$\Xi^- \Lambda$ &  $-0.053(10)$  &$-0.06(4)$    \\
$\Xi^- \Sigma^0$ & $-0.068(17)$    &$-0.05(2)$         \\
\hline
\end{tabular}
\caption{SU(3) breaking corrections to the $\Delta S=1$ vector charges.  The LQCD results are from Ref. \cite{Shanahan:2015dka}. }
\label{table:df1}
\end{center}
\end{table}

For   the charge radii the loop contributions are from diagrams $A_3$ and $B_2$ and the renormalization  is provided by  the LECs $g_1$ and $g_2$ in ${\cal{L}}^{(3)}_B$ and ${\cal{L}}^{(4)}_B$ respectively, of which only $g_1$ is required for canceling the loop UV divergence according to Eqn. (\ref{eqn:chargeFFUV}) \footnote{In Ref. \cite{Fernando:2017yqd} the finite term proportional to $g_2$  was  overlooked. }.  As is the case with form factors in ChPT, the charge radii depend  logarithmically in the GB masses. They can be determined by fitting to the known electric charge radii of proton, neutron  and $\Sigma^-$, or simply fixed using the first two. If one wishes to study also the large $N_c$ limit, an assignment at generic $N_c$ of the quark electric charges has to be done. One such an assignment that  respects all gauge and gauge-gravitational anomaly cancellations in the Standard Model is    is given by \cite{Shrock:1995bp} $\hat Q=T^3+\frac{1}{\sqrt{3}}T^8+\frac{3-N_c}{6N_c} B$.   The last term comes from the baryon number charge $B$, and can be implemented by simply adding to the Lagrangians the corresponding terms with an $SU(3)$ singlet vector source field.  This charge operator gives for the states   identified with the physical octet and decuplet the same electric charges as the physical ones for any $N_c$. The analysis of the charge radii in the present framework is revealing: in the strict large $N_c$ limit one finds that the non-analytic loop contributions to the $T^3$ charge radius of nucleons by Diagram  $A_3$ is $\ord{N_c^0}$, where the contribution is driven by the hyperfine mass splitting term, i.e, for $C_{HF}\to 0$ the contribution becomes $\ord{1/N_c}$, and Diagram $B_2$ gives only contributions $\ord{1/N_c}$. For the charge $T^8$ the loop contributions are $\ord{N_c^0}$. One however notes that for the physical $\pi$ and $K$  meson masses the non-analytic terms join the large $N_c$ scaling at rather large $N_c$.
 The charge radii of the neutral baryons receive only UV finite loop contributions and are renormalized only by the finite $g_2$ term.

Using the three known charge radii, $g_{1,2}$ are determined modulo the main uncertainty  stemming from the value  used for $\gA$. At the renormalization scale $\mu=m_\rho$, using the value of $\gA\sim 1$ obtained by the analysis of the axial couplings \cite{Fernando:2017yqd},  $C_{HF}\sim 200$ MeV,  and with $\Lambda=m_\rho$ one finds $g_1\simeq 1.33$ and $g_2\simeq 0.74$.   $g_2$ is sensitive to $C_{HF}$, which is understood as a result that the non-analytic contributions to the neutron radius is very important, and thus sensitive to that parameter, while $g_1$ is not.  One also observes that  both LECs are crucial for obtaining a good description of the radii. For the used value of $\mu$, the fraction of the loop contribution to $\langle r^2\rangle$ of the proton is 15\%, and for the neutron  it is about 60\%. The short distance contributions are thus very important in both cases. 
The dominant non-analytic contributions to the radii are  proportional to $\log m_q$, with other non-analytic terms involving the LEC $C_{HF}$ giving almost negligible contributions, making the results insensitive to it.
Table \ref{table:chargerms} shows the results for the charge radii of the baryon octet along with the contributions by the CTs. The latter contributions to $\langle r^2\rangle$ satisfy the exact linear relation, in obvious notation: $a \Lambda+p+\Sigma^++\frac 13(a-4) (n+\Sigma^0+\Xi^0)+\Sigma^-+\Xi^-=0$ valid for any $a$ and resulting from the electric charge being a U-spin singlet; it is violated only by finite $SU(3)$ breaking loop contributions. The isotriplet nucleon charge radius is $\ord{N_c^0}$, while the isosinglet one receives  loop and $g_2$ contributions $\ord{N_c^0}$ and a $g_1$ contribution   $\ord{N_c}$, where the $\ord{N_c}$ term contribution to the EM charge radius must be cancelled by adding to the Lagrangian a finite charge-radius CT proportional to baryon number and weighted according to the electric charge assignment at arbitrary $N_c$ mentioned above.

\begin{table}[h!]
 \begin{tabular}{lccc}
\hline
&&$\langle r^2 \rangle {\rm[fm^2]}$ &\\
	 	& Full   & CT & Exp\\ \hline
p		& 0.707 & 0.596   & $0.7071(7)$\\ 
%	\hline 
	n & $-0.116$   &$ -0.049$& $-0.116(2)$ \\ 
%	\hline 
	$\Lambda$ & $-0.029$  & $-0.024$& $\cdots$ \\ 
%	\hline 
	$\Sigma^+$ & 0.742  & 0.596& $\cdots$ \\ 
%	\hline 
	$\Sigma^0$ & 0.029  &  0.024& $\cdots$ \\ 
%	\hline 
	$\Sigma^-$ & 0.683  & 0.548 & $0.608(156)$  \\ 
%	\hline 
	$\Xi^0$ & $-0.016$   & $-0.049$& $\cdots$ \\ 
%	\hline 
	$\Xi^-$ & 0.633  & 0.548& $\cdots$ \\ 
	\hline  
\end{tabular} 
\caption{Electric charge radii of octet baryons.  The proton and neutron radii  are inputs.  The proton radius used is the one resulting from the muonic Hydrogen Lamb shift \cite{PhysRevD.98.030001}.
 The second column shows the contribution by contact terms $g_{1,2}$  for $\mu=m_\rho$.}
 \label{table:chargerms}
\end{table}
 
At the present order in the $\xi$ expansion, the curvature of the form factors, proportional to $\langle r^4\rangle=60 \frac{d^2 f_1}{d (Q^2)^2}$, is given by the one-loop non-analytic terms with contributions that are inversely proportional to quark masses.  The curvature is nominally an effect $\ord{\xi^4}$ in the form factor, which therefore receives contributions from terms $\ord{\xi^6}$ in the Lagrangian, and only in the limit of sufficiently small quark masses will the non-analytic contributions obtained here be dominant.   In the recent work of Ref. \cite{Alarcon:2018irp} the electric charge    higher moments have been studied, where    t-channel elastic unitarity has been implemented in the EFT along with the constraints of the $1/N_c$ expansion  \cite{Granados:2013moa,Granados:2016jjl,Alarcon:2017asr,Alarcon:2017lhg,Alarcon:2018irp}. In particular,  for the curvature  they find  $\langle r^4\rangle^p=0.735(35)\, {\rm fm^4}$ and  $\langle r^4\rangle^n=-0.540(35) \,{\rm fm^4}$, to be compared with the one-loop contributions found here, 0.032 and $-0.021\, {\rm fm^4}$ respectively,  roughly a factor 25 smaller in magnitude in each case. Clearly the description of the curvature must be primarily given by higher order contact terms, and 
   to the order of the expansion followed here,  the failure to account for the curvature  limits the  present description  of charge form factors  to  the expected range  given by the  radii,  $Q^2\lesssim 0.05 ~{\rm GeV ^2}$. 
\section{Magnetic moments}
\label{sec:magneticmoments}
As mentioned earlier, at lowest order the magnetic moments of all vector currents are given in terms of the single LEC $\kappa$.  In particular, using the EM current        the LO value of $\frac\kappa\Lambda$   can be fixed from   the proton's magnetic moment $\mu_p$ in units of  the nuclear magneton $\mu_N$, namely    
$e\,\frac{\kappa}{2\Lambda} =\mu_p=2.7928\; \mu_N$. Also, the $M_1$ radiative decay width of the $\Delta$  at LO is given by:
\beq
\Gamma_{\Delta\to N\gamma}=\frac{e^2}{9 \pi}\left( \frac{\kappa}{\Lambda}\right)^2 \frac{m_N}{m_{\Delta} }\omega^3,
\label{eqn:deltaEMwidth}
\eeq 
where $\omega$ is the photon energy.  Using the above result for $\frac\kappa\Lambda$    gives  $\Gamma^{\rm LO}_{\Delta\to N\gamma}=0.38\;{\rm MeV} $, to be compared with the experimental value $0.70\pm 0.06$ MeV. In terms of the transition magnetic moment, the LO result is $\mu_{\Delta^+ p}=\frac{2\sqrt{2}}{3}\,\mu_p$ while the experimental one from Eqn.(\ref{eqn:deltaEMwidth})  and from the helicity $N-\Delta$ photo-couplings  \cite{Tiator:2000iy} are  $3.58(10)\mu_N$   and $3.46(3)\mu_N$ respectively. This shows the need for a significant spin-symmetry breaking effect of 30\% to be accounted for by the higher order corrections.

 The LO magnetic moment  operator    $G^{ia}$ is proportional to the LO axial currents, and the NLO effects   stem from  quark masses and   spin symmetry breaking. In the strict large $N_c$ limit those corrections scale as follows: $SU(3)$ breaking corrections $\ord{(m_s-\hat m) N_c}$, i.e. the same scaling in $N_c$ as the LO term, and spin symmetry breaking corrections $\ord{1/N_c}$, i.e. $\ord{1/N_c^2}$ with respect to the LO term, well known from tree level analyses in Refs.   \cite{Dashen:1993ac,Jenkins:1994md}.

 The experimentally available magnetic moment  ratios and the corresponding LO results are represented in Table \ref{table:magmomratios}.
\begin{table}[h]
 \begin{center}
\begin{tabular}{lcc}
\hline
~	 ~& ~~~Exp~~~ & ~~~LO~~~\\
	\hline
$p/n$ & $ -1.46 $ & $ -1.5 $ \\ 
$\Sigma^+/\Sigma^- $ & $ -2.12 $ & $ -3 $ \\ 
	$\Lambda/\Sigma^+ $ & $ -0.25 $ & $-\frac13$ \\ 
	$p/\Sigma^+$ & $ 1.14 $ & $ 1 $ \\ 
	$\Xi^0/\Xi^- $ & $ 1.92 $ & $ 2 $ \\ 
	$p/\Xi^0 $ & $ -2.23 $ & $ -1.5 $ \\ 
	$\Delta^{++}/\Delta^+ $ & $ 1.4(2.8) $ & $ 2 $ \\ 
	$\Omega^-/\Delta^+$ & $ -0.75 $ & $ -1 $ \\ 
	$p/\Delta^+$ & $ 1.03 $ & $ 1 $ \\
	$p/(\Delta^+ p)$ & $ 0.78$ & $ \frac{3}{2\sqrt{2}} $ \\ 
	$p/(\Sigma^{*0}\Lambda)$ & $ 1.02$ & $ \sqrt{\frac{3}{2}} $ \\
	$p/(\Sigma^{*+}\Sigma^+)$ & $ -0.88$ & $ -\frac{3}{2\sqrt{2}} $ \\
	\hline
\end{tabular}
\caption{LO ratios of magnetic moments. }
\label{table:magmomratios}
\end{center}
\end{table}
 It is evident that there are significant $SU(3)$ breaking effects, which together with the important spin-symmetry breaking observed. in particular   in the $\Delta N$  $\rm M_1$ amplitude indicate the relevance of the NNLO calculation. Note that   all weak magnetic moments, i.e., magnetic moments associated with the $\Delta S=1$ currents are also fixed at LO,  as they are empirically unknown. In the case of the neutron $\beta$  decay the weak magnetic term is obtained from the isovector part of the EM magnetic moments of proton and neutron, which in this case, due to isospin symmetry, is quite accurate. On the other hand, in hyperon beta decay the effect of weak magnetism is too small to be at present experimentally accessible. Fortunately the advent of LQCD calculations of magnetic moments  with increasing accuracy will allow the study of weak magnetism.

The one loop corrections to the magnetic moments are obtained from the spatial components of the vector currents depicted in Fig. \ref{fig:1-loop-VC}, where the contributions stem from diagrams $A$ and $B_1$. Diagrams $A_{1,2}$ involve  $\Gamma\propto G^{ia}$, which is similar to the  axial currents already analyzed in Ref. \cite{Fernando:2017yqd}. The loop contributions to the $Q^2$ dependence of the magnetic form factors  stem  from  diagram  $A_3$. 

The UV divergencies of the one loop diagrams contributing to the magnetic moments after reduction of the corresponding expressions Eqns.(\ref{eqn:A12UV})  and  (\ref{eqn:A3B12UV}) using a basis of spin-flavor operators read as follows:
\bea
V^{\mu a}_{Mag}(A_{1+2})^{UV}&=& i\frac{\lambda_\eps}{(4\pi)^2} \frac{\kappa}{2\Lambda} \left(\frac{\mathring g_A}{F_\pi} \right)^2\eps^{ijk} q^j 
\left(-B_0 \Big( \frac{23}{6} m^0 G^{ka}+\frac{11}{24} d^{abc} m^b G^{kc}+\frac{5}{18} m^a S^k\Big)\right.\nonumber\\
&+&\left. \frac 23 \left( \frac{C_{HF}}{N_c}\right)^2 \Big((N_c(N_c+6)-3) G^{ka}+8 \{\hat S^2,G^{ka}\}+8 S^k S^m G^{ma}-\frac{11}{2}(N_c+3) S^k T^a\Big)\right)\nonumber\\
V^{\mu a}_{Mag}(A_3)^{UV}&=& i\frac{\lambda_\eps}{(4\pi)^2} \left(\frac{\mathring g_A}{F_\pi} \right)^2 \frac{C_{HF}}{N_c} \eps^{ijk} q^j \left(\frac{N_c+3}{2} G^{ka} -2 S^kT^a  \right)\nonumber\\
V^{\mu a}_{Mag}(B_1)^{UV}&=&- i\frac{\lambda_\eps}{(4\pi)^2} \frac{\kappa}{2\Lambda}  \frac{1}{F_\pi^2}  \eps^{ijk} q^j B_0\left(6\,m^0 G^{ka}+\frac 32 d^{abc}m^b G^{kc} \right),
\eea
adding up to:
\bea
V_{Mag}^{UV^{\mu a}}&=&\frac{i \lambda _\epsilon  q^j \eps^{ ijk} }{16 \pi ^2 {F_\pi }^2 {\Lambda }}\left(-\frac{1}{12}\kappa {B_0} \left( 
   (\frac{11}{4} {\gA}^2+9)  m^b {G}^{kc}
   {d }^{abc}+  \left(23 {\gA}^2+36\right){m^0}
   {G}^{ka}+\frac{5}{3} {\gA}^2  m^a
   {S}^k\right) \right.
   \nonumber\\
   &+& \frac{{C_{HF}} {\gA}^2 }{6
   {N_c}^2} \left(2\,  \kappa  \,{C_{HF}}
   (({N_c} ({N_c}+6)-3){G}^{ka}+8 \{\hat S^2,{G}^{ka}\})+3 {\Lambda} {N_c} ({N_c}+3){G}^{ka}\right.
 \nonumber\\
 &+& \left. \left. 16\,
 \kappa\,  {S^m}{G}^{ma} {S}^k-{S^k}{T}^{a} (11\, \kappa \, {C_{HF}}
 ({N_c}+3)+12 {\Lambda  } {N_c})   \right)\right)
 \label{eqn:UVMagMom}
\eea

The renormalization of the  magnetic moments  is provided by the Lagrangians with the LECs $\kappa_{D,F,1,\cdots,5}$, and the magnetic radii receive  only finite one-loop contributions and a finite renormalization by the term $\kappa_r$. The $\beta$ functions  of the magnetic LECs resulting from Eqn.(\ref{eqn:UVMagMom}) are shown  in Table \ref{table:magbetaf}.
\begin{table}[h]
 \begin{center}
 \begin{tabular}{ll}
 \hline
 $LEC$   ~~~~~& $\beta \times F_\pi^2$\\
 \hline
 $\kappa $ &  $\Lambda\, \gA^2\, \frac{C_{HF}}{N_c} \left(\frac 12 (N_c+3)+\frac 13 (N_c(N_c+6)-3)\frac{\kappa}{\Lambda}\frac{C_{HF}}{N_c} \right) $          \\
 $\kappa_1 $ &    $ -\Lambda\, \gA^2\, C_{HF}\left(2+\frac{11}{6}(N_c+3) \frac{\kappa}{\Lambda}\frac{C_{HF}}{N_c}\right)$          \\
$\kappa_2 $ & $-\Lambda^2 \,\kappa \,\left(3+\frac{23}{12}\,\gA^2\right)$             \\
$\kappa_D $ & $-\Lambda^2 \,\kappa \,\left(\frac{3}{4}+\frac{11}{48}\,\gA^2\right)$             \\
$\kappa_F $ & $0$             \\
$\kappa_3 $ &   $-\Lambda^2 \,\kappa \,\frac{5}{36} \, \gA^2$         \\
$\kappa_4 $ & $\frac{8}  {3} \gA^2 \,\kappa \,C_{HF}^2$           \\
$\kappa_5 $ &  $\frac{8}  {3} \gA^2 \,\kappa \,C_{HF}^2$                 \\
$\kappa_r$ & 0 \\	\hline
\end{tabular}
\caption{$\beta$ functions of LECs associated with magnetic moments and  radii. The renormalized LECs are defined according to $X=X(\mu)+\frac{\beta_X }{(4\pi)^2} \lambda_\eps$.}
\label{table:magbetaf}
\end{center}
\end{table}

For $N_c=3$ the set of local terms that contribute to the magnetic moments remains linearly independent. If one only considers the EM current, the term proportional to $\kappa_F$ does not contribute, and for the known magnetic moments together with the information on the   $M_1$ transition $\Delta\to N\gamma$  one can fit the rest of the LECs. Note that in the absence of information on the $SU(3)$ singlet quark mass $m^0$ dependence, the LEC $\kappa_2$ is subsumed into $\kappa$, and  the lack of knowledge on the $\Delta S=1$ weak magnetic moments does prevents at present  a determination of $\kappa_F$.

The results of the fits are shown in Table \ref{table:magmom}. Since the input magnetic moments have errors (much) smaller than the theoretical error of the present calculation estimated to be of the order of NNNLO corrections or about 5\%,  the $\chi^2$ has been normalized for  estimating  the LECs' errors. Important correlation is found between the following pairs of LECs: $\kappa_4-\kappa_5$, $\kappa_4-\kappa_6$ and $\kappa_5-\kappa_6$.

\begin{table}[htp]
\begin{center}
\begin{tabular}{ccc}
\begin{tabular}{cll}
\hline  
 LEC$\times \frac{m_N}{\Lambda}$  & LO & NNLO\\
\hline
$\kappa$  & 2.80 & 2.87(2) \\
 {$\kappa _1$} & 0 & 3.18 (10)\\
 {$\kappa _2$} & 0 & 0. \\
 {$\kappa _D$} & 0 & 0.46 (5)\\
 {$\kappa _F$} &0 & $\cdots$ \\
 {$\kappa _3$} & 0 & 0.51(6) \\
 {$\kappa _4$} & 0 & $-2.84(40)$ \\
 {$\kappa _5$} & 0 & 1.19(20) \\
\hline
\end{tabular}
&\,
\begin{tabular}{lccc}\hline
 { } & $\mu _{LO}$  &$ \mu _{NNLO}$& $\mu
   _{{Exp}}$ \\ \hline
 {p} & 2.691 & 2.797 & $2.7928(23)$ \\
 {n} & $-1.794$ &$ -1.929$ &$ -1.9130(45)$ \\
 {$\Sigma^ +$} & 2.691 & 2.359 & $2.46(1)$ \\
 {$\Sigma ^0$} & 0.897 & 0.834 & $\cdots$ \\
 {$\Sigma ^-$} & $-0.897$ & $-0.691$ & $-1.16(3)$ \\
$ \Lambda$  & $-0.897$ &$ -0.595 $& $-0.613(4)$ \\
 {$\Xi ^0$} & $-1.794 $& $-1.245 $& $-1.250(14)$ \\
 {$\Xi ^-$} & $-0.897$ & $-0.657$ & $-0.6507(25)$ \\
 $\Delta^+  p$ & 2.537 & 3.580 & $3.58(10)$  \\
 $\Sigma^0 \Lambda $ & 1.553 & 1.562 & $1.61(8)$  \\
 $\Sigma^{*0} \Lambda $ & 2.197 &  2.685 & $2.73(25)^{\rm a}$  \\
 $\Sigma^{*+} \Sigma^+ $ & $-2.537$ & $-2.326$  & $-3.17(36)^{\rm b}$  \\
 \hline
\end{tabular}
&\,
\begin{tabular}{lccc}\hline
 { } &$ \mu _{{LO}} $ & $\mu _{{NNLO}} $ & $\mu
   _{{Exp}} $\\
\hline
 {$\Delta ^{++}$} & 5.381 & 5.979 & $3.7 - 7.5$\\
$ \Delta^+$  & 2.691 & 3.027 & $2.7(1.2) $\\
 {$\Delta ^0$} & 0 & 0.074 & $\cdots$ \\
 {$\Delta ^-$} & $-2.691$ & $-2.879$ &$\cdots$ \\
 {$\Sigma^{*+}$} & 2.691 & 3.163 & $\cdots$ \\
 {$\Sigma ^{*0}$} & 0 & 0.315 & $\cdots$ \\
 {$\Sigma ^{*-}$} &$ -2.691$ & $-2.534$ & $\cdots$ \\
 {$\Xi ^{*0}$} & 0 & 0.496 &$\cdots$ \\
 {$\Xi ^{*-}$} & $-2.691$ &$ -2.242 $& $\cdots$ \\
$ \Omega$  & $-2.691$ & $-2.005$ & $-2.02(5)$ \\
 \hline
\end{tabular}
\end{tabular}
 \caption{Results from fits to the electric current magnetic moments, in units of the  nuclear magneton $\mu_N$.
The  renormalization scale was set to $\mu=\Lambda=m_\rho$. $\kappa_F$ requires $\Delta S=1$ weak magnetic moments to be determined. Empirical results from PDG and  references $^{\rm a}$\cite{Keller:2011nt},   $^{\rm b} $\cite{Keller:2011aw}.}
\label{table:magmom}
\end{center}
\end{table}%

As mentioned earlier, the   $\Delta N \gamma$ amplitude at LO is too small by roughly 30\%, a manifestation of an important spin-symmetry breaking effect.  The effect receives a small non-analytic contribution (at $\mu=m_\rho$), and the contributions from the contact terms are as follows:  $\kappa_D: \ord{(m_s-\hat m)N_c}$, and $\kappa_4: \ord{1/N_c}$. From the fit one finds a modest contribution from   $\kappa_D$ and a dominant contribution from   $\kappa_4$. Since the latter is a $1/N_c^2$ correction with respect to the LO magnetic moment, it seems  to be unnaturally large. This is a bit surprising as a similar kind of effect  in the $\Delta N$  axial vector coupling is actually unnaturally  small. This contrast remains to be understood.    Finally, a fit where the $\Delta N $ transition is not an input shows   an enhancement   but only by about half of what is needed.

An interesting case is the magnetic moment of $\Sigma^{*0}$: all LO and NLO tree level and quark mass independent contributions vanish, receiving only NNLO tree and loop contributions which vanish in the SU(3) symmetry limit.  On the other hand, the experimental value of the magnetic moment of   $\Sigma^-$ quoted as average by the PDG \cite{PhysRevD.98.030001} cannot be described: U-spin symmetry implies that it must be equal to the magnetic moment of the $\Xi^-$ up to NNLO $SU(3)$ breaking by quark masses. The experimental results imply a very large  effect
which is very difficult to reconcile with the other U-spin multiplets, where the effect is between 12\%  and 25\%  per unit of strangeness, while for the pair $\Sigma^-$ $\Xi^-$ case it is 44\%!.
  
 One of the early tests of the magnetic moments in $SU(3)$ was provided by the Coleman-Glashow (CG) relation, namely $\mu_p-   \mu_n -  \mu_{\Sigma^+}+   \mu_{\Sigma^-}  + \mu_{\Xi^0}  - \mu_{\Xi^-}   =0$. This relation remains valid at tree level NNLO and receives only a finite correction from the one loop contributions. Explicit calculation gives the deviation with estimated theoretical error $\Delta_{CG}=1.09\pm 0.25\,\mu_N$ to be compared with the experimental deviation $0.49\pm 0.03\,\mu_N$, affected however by the $\Sigma^-$ issue.
  
 Finally, the weak interaction magnetic moments for hyperon decays turn out to depend on the LEC $\kappa_F$ which does not appear in the EM case.  The result for the LECs from the EM case gives the predictions: $\mu_{\Sigma^- n}= (0.516-0.180 \;\kappa_F) \frac{g}{2m_N}$ and
$\mu_{\Lambda p}= (-1.41+0.66  \; \kappa_F) \frac{g}{2m_N} $, where $g=e/\sin \theta_W$. At LO one has the large hierarchy  $\mu_{\Lambda p}/\mu_{\Sigma^- n}=-\sqrt{27/2}$. A determination of $\kappa_F$ will require a LQCD calculation.

  \subsection{Magnetic radii}
  The magnetic radii are theoretically very constrained at the order of the present calculation. For all the vector currents and baryons they are determined only by UV finite loop contributions and the single available finite counterterm fixed by the LEC 
   $\kappa_r$.  Since only the magnetic radii of proton and neutron are experimentally known, one can use these to fit that LEC  leading to  the results shown in Table \ref{table:magradii}.
   The rest of the radii are then predictions which can hopefully be tested in the future with LQCD calculations. Note that the lion share of the magnetic radii is from  the short distance terms proportional to $\kappa_r$ with the loop contribution from diagram $A_3$ in Fig. \ref{fig:1-loop-VC} giving up to 20\% for proton, neutron and $\Sigma^-$ and less than 10\% for the rest.
\begin{table}[h]
 \begin{center}
\begin{tabular}{llll}
\hline  $\kappa_r=-2.63$ ~~~~&&$\langle r^2 \rangle {\rm[fm^2]}$ & \\
& 
  $ {\rm Exp}  $ & 
  $ ~~{\rm Th}  $& 
  $ {\rm Loop} $\\
  \hline
  {p } &  0.724 & 0.718 & 0.134 \\
 {n } & 0.746
 & 0.747 & 0.179 \\
 {$\Sigma ^+$} & $\cdots$ & 0.766 & 0.100 \\
 {$\Sigma ^0$} & $\cdots$ & 0.698 & 0.061 \\
 {$\Sigma ^-$} & $\cdots$ & 0.922 & 0.189 \\
 {$\Lambda $} & $\cdots$ & 0.895 & 0.079 \\
 {$\Xi^0 $} & $\cdots$ & 0.872 & 0.081 \\
 {$\Xi ^-$} & $\cdots$ & 0.796 & 0.035 \\
 $\Delta^+ p$& $\cdots$ & 0.875 & 0.226\\
  \hline	
\end{tabular} \caption{Magnetic radii  from a fit to  nucleons.}
\label{table:magradii}
\end{center}	
\end{table}

Finally, a calculation of the curvature of the EM magnetic moments yields: $\langle r^4\rangle^p=0.38\, {\rm fm^4}$ and  $\langle r^4\rangle^n=0.54 \,{\rm fm^4}$ to be compared with those obtained in   Ref. \cite{Alarcon:2018irp}, which are respectively $1.72(6)$ and $2.04(1) \,{\rm fm^4}$, leading to  a similar assessment as in the case of the electric charge already discussed, although less dramatic.

\section{Summary}
\label{sec:Conclusions}

This work presented the study of the $SU(3)$ vector currents in baryons based on  the combined chiral and $1/N_c$ expansion. It was carried out in the context of the $\xi$ power counting   to one-loop. This corresponds to a calculation of the charges, magnetic moments and their radii for both octet and decuplet baryons. The calculations have been provided for generic $N_c$, which permits an exploration of the behavior of those observables with respect to the number of colors.  Only two LECs are needed to determine all $SU(3)$ charge radii, while the magnetic moments need to be renormalized involving eight LECs, of which all but two can be fixed solely in terms of the known EM magnetic moments. Of the two remaining LECs, one needs information about $\Delta S=1$ weak magnetic moments and the second requires knowledge of magnetic moments at different values of quark masses, which can be obtained from LQCD calculations. Finally the magnetic radii are all determined in terms of a single LEC.  
The fits indicate that the values   LECs are   within the range of  natural magnitude, although there is a puzzling issue, namely  the unnaturally large spin-symmetry breaking required for the description of the $\Delta N$ transition magnetic moment.  Finally, the curvature of form factors    is given at the order of the calculation by  non-analytic terms in $m_q$, which  turn out to be very small, and therefore requiring for their description an extension of  the present work to higher order.

\begin{acknowledgments}
The authors thank  Rub\'en Flores Mendieta and   Christian Weiss   for very useful discussions. 
This work was supported by DOE Contract No. DE-AC05-06OR23177 under which JSA operates the Thomas Jefferson National Accelerator Facility, and by the National Science Foundation  through grants PHY-1307413 and PHY-1613951. 
\end{acknowledgments}
\newpage

\appendix
 
\section{Spin-flavor algebra }
\label{app:Algebra}

The $4N_f^2-1$ generators of the spin-flavor group $SU(2 N_f)$  consist of the three spin generators $S^i$, the $N_f^2-1$   flavor $SU(N_f)$ generators  $T^a$, and the remaining $3(N_f^2-1)$   spin-flavor generators $G^{ia}$. The commutation relations are:
\bea
~& [ S^i,S^j ]= i \eps^{ijk}S^k,~~ [ T^a,T^b ]=i f^{abc} T^c ,~~ [ T^a,S^i ]=0& ,\nonumber\\    
~& [ S^i,G^{ja} ]=i  \eps^{ijk} G^{ka},~~ [T^a,G^{ib}]=i f^{abc} G^{ic}&,\nonumber\\   
~ & [ G^{ia},G^{jb} ]= \frac{i}{4}\delta^{ij}f^{abc}T^c+\frac{i}{2 N_f}\delta^{ab} \eps^{ijk} S^k+\frac{i}{2}\eps^{ijk}d^{abc}G^{kc}&.
\label{eq:commutation-relations}
\eea
 
In spin-flavor representations with $N_c$ indices corresponding to baryons, the generators $G^{ia}$ have matrix elements $\ord{N_c}$ on states with $S=\ord{N_c^0}$.  
The  ground state baryons furnish the totally symmetric irreducible representation of $SU(6)$ with $N_c$ Young boxes, which decomposes into the following $SU(2)_{\text{spin}}\times SU(3)$ irreducible representations: $[S,(p,q)]=[S,(2S,\frac{1}{2}(N_c-2S))]$, $S=1/2,\cdots,N_c/2$ (assumed $N_c$ is odd). The baryon states can  then be denoted by: $\mid \! S S_3,YII_3\rangle$, where the spin $S$ of the baryon determines its $SU(3)$ multiplet.

\subsection{Matrix elements of the SU(6)  generators} 
In general the matrix elements of a $SU(2)_{\text{spin}}\times SU(3)\subset SU(6)$ tensor operator between baryons   ground state baryons are given by the Wigner-Eckart theorem, with obvious notation:
\bea
\langle S'S'_3,R'\;Y'I'I'_3\mid O^{\ell \ell_3}_{\tilde R \tilde Y\tilde I\tilde I_3}\mid \! SS_3,R\;YII_3\rangle &=&\frac{1}{\sqrt{2S'+1}\sqrt{\text{dim}R'}}\;\langle SS_3,\ell\ell_3\mid \! S'S'_3\rangle \\
	\times  \sum_{\gamma } \langle S',R'\mid\mid O^{\ell }_{\tilde R  }\mid\mid \! S,R\rangle_\gamma&&\!\!\!\!
	\left\langle
	\begin{array}{cc}
		R  & \tilde R       \\
		Y ~I ~I_{3}     & \tilde Y~\tilde I~ \tilde I_3
	\end{array}
	\right|   \left.
	\begin{array}{c}
		 R'          \\
Y' ~ I' ~ I'_3
	\end{array} \right\rangle_\gamma  ~~,\nonumber
\eea
where $R$ represents the $SU(3)$ multiplet of the baryon, and $\gamma$ indicates the  possible recouplings in $SU(3)$.  The matrix elements of interest are then given by:
\bea
\langle S'S'_3,Y'I'I'_3\mid \! S^m\mid \! SS_3,YII_3\rangle&=& \delta_{SS'}\delta_{YY'}\delta_{II'}\delta_{I_3I'_3}\sqrt{S(S+1)} \langle SS_3,1m\mid \! S'S'_3\rangle\nonumber\\
\langle S'S'_3,Y'I'I'_3\mid T^{yii_3}\mid \! SS_3,YII_3\rangle&=&\delta_{SS'}\delta_{S_3S'_3}\frac{1}{\sqrt{\text{dim}(2S,\frac 12(N_c-2S))}}\langle S\mid\mid T\mid\mid \! S\rangle\nonumber\\
&\times& 	\left\langle
\begin{array}{cc}
	(2S,\frac 12(N_c-2S))  & (1,1)      \\
	Y ~I ~I_{3}     & yii_3
\end{array}
\right|   \left.
\begin{array}{c}
(2S,\frac 12(N_c-2S))          \\
	Y' ~ I' ~ I'_3
\end{array} \right\rangle_{\gamma=1}\nonumber\\
\langle S'S'_3,Y'I'I'_3\mid G^{m,yii_3}\mid \! SS_3,YII_3\rangle&=&
\frac{\langle SS_3,1m\mid \! S'S'_3\rangle}{\sqrt{2S'+1}\sqrt{\text{dim}(2S,\frac 12(N_c-2S))}}\\
\times \sum_{\gamma=1,2 } \langle S'\mid\mid G\mid\mid \! S\rangle_\gamma &&\!\!\!\!	\left\langle
\begin{array}{cc}
	(2S,\frac 12(N_c-2S))  & (1,1)      \\
	Y ~I ~I_{3}     & yii_3
\end{array}
\right|   \left.
\begin{array}{c}
	(2S,\frac 12(N_c-2S))          \\
	Y' ~ I' ~ I'_3
\end{array} \right\rangle_{\gamma}\nonumber
\eea
where the reduced matrix elements are (here $p=2S$, $q=\frac 12(N_c-2S)$):
\small{
\bea
\!\!\!\!\!\!
\langle S\mid\mid T\mid\mid \! S\rangle&=& \sqrt{\text{dim}(	p,q) C_2(	p,q)}\nonumber\\
= &&\!\!\!\!\!\frac{\sqrt{(2 S+1) ({N_c}-2
		S+2) ({N_c}+2 S+4)
		({N_c} ({N_c}+6)+12 S
		(S+1))}}{4 \sqrt{6}}\nonumber\\
\langle S'\mid\mid G\mid\mid \! S\rangle_{\gamma=1}\!\!&=&\!\!\left\{\begin{array}{ll}
\text{if~~}S=S'+1:&-\frac{\sqrt{\left(4 S^2-1\right)
		\left(({N_c}+2)^2-4
		S^2\right)
		\left(({N_c}+4)^2-4
		S^2\right)}}{8 \sqrt{2}}\\
\text{if~~}S=S'-1:&-\frac{\sqrt{(4 S (S+2)+3)
		({N_c}-2 S) ({N_c}-2
		S+2) ({N_c}+2 S+4)
		({N_c}+2 S+6)}}{8 \sqrt{2}}\\
\text{if~~}S=S':&\text{sign}(N_c-2S-0^+)\frac{({N_c}+3) (2 S+1)
	\sqrt{S (S+1) ({N_c}-2 S+2)
		({N_c}+2 S+4)}}{ 
	\sqrt{6{N_c}
		({N_c}+6)+12 S (S+1)}}
 \end{array} \right. \\
\langle S'\mid\mid G\mid\mid \! S\rangle_{\gamma=2}&=&{\footnotesize -\delta_{SS'}\frac{(2 S+1) \sqrt{({N_c}-2
		S) ({N_c}+2 S+6)
		\left(({N_c}+2)^2-4
		S^2\right)
		\left(({N_c}+4)^2-4
		S^2\right)}}{8 \sqrt{2}
	\sqrt{{N_c}
		({N_c}+6)+12 S (S+1)}}}  \nonumber
\eea}

\section{Loop integrals}
\label{app:loopintegrals}
The one-loop integrals needed in this work are provided here.  The definition  $\widetilde{d^dk}\equiv d^d k/(2\pi)^d$ is used.

The scalar and tensor one-loop integrals are:
\bea
I(n,\alpha,\Lambda)\equiv\int \widetilde{d^dk}\; \frac{k^{2n}}{(k^2-\Lambda^2)^\alpha}&=&i(-1)^{n-\alpha} \frac{1}{(4\pi)^{\frac {d}{2}}} \frac{\Gamma(n+\frac{d}{2})\Gamma(\alpha-n-\frac{d}{2})}{\Gamma(\frac{d}{2})\Gamma(\alpha)}\left(\Lambda^2\right)^{n-\alpha+\frac{d}{2}}\nonumber\\
I^{\mu_1,\cdots,\mu_{2n}}(\alpha,\Lambda)\equiv\int \widetilde{d^dk}\; \frac{k_{\mu_1}\cdots k_{\mu_{2n}}}{(k^2-\Lambda^2)^\alpha}&=&i(-1)^{n-\alpha} \frac{1}{(4\pi)^{\frac{d}{2}}}\frac{1}{4^n n!} \frac{ \Gamma(\alpha-n-\frac{d}{2})}{ \Gamma(\alpha)}\left(\Lambda^2\right)^{n-\alpha+\frac{d}{2}}\nonumber\\
&\times& \sum_{\sigma} g_{\mu_{\sigma_1}\mu_{\sigma_2}}\cdots g_{\mu_{\sigma_{2n-1}}\mu_{\sigma_{2n}}} \\
&=&\frac{1}{4^n n!}\frac{\Gamma(\frac d2)}{\Gamma(n+\frac d2)} I(n,\alpha,\Lambda) \sum_{\sigma} g_{\mu_{\sigma_1}\mu_{\sigma_2}}\cdots g_{\mu_{\sigma_{2n-1}}\mu_{\sigma_{2n}}}~~,\nonumber
\eea
where $\sigma$ are the permutations of $\{1,\cdots,2n\}$.

The Feynman parametrizations needed when heavy propagators are in the loop are as follows:
\bea
\frac{1}{A_1\cdots A_m B_1\cdots B_n}&=& 2^m \Gamma(m+n) \int_0^\infty d\lambda_1\cdots d\lambda_m\int_0^1 d\alpha_1\cdots d\alpha_n \delta(1-\alpha_1-\cdots-\alpha_n)\nonumber\\
&\times& \frac{1}{(2\lambda_1 A_1+\cdots+2\lambda_m A_m+\alpha_1 B_1+\cdots+\alpha_n B_n)^{m+n}},
\eea
where the $A_i$ are heavy particle static propagators denominators, and the $B_i$ are relativistic ones.

The integration over a Feynman parameter $\lambda$ is of the general form:
\beq
J(C_0,C_1,\lambda_0,d,\nu)\equiv \int_0^\infty (C_0+C_1(\lambda-\lambda_0)^2)^{-\nu+\frac{d}{2}} d\lambda,
\eeq
which satisfies the recurrence relation:
\bea
J(C_0,C_1,\lambda_0,d,\nu)&=&\frac{-\lambda_0  (C_0+C_1\lambda_0^2)^{1-\nu+\frac{d}{2}}+(3+d-2\nu) J(C_0,C_1,\lambda_0,d,\nu-1)}{(d-2\nu+2)C_0}\nonumber\\
J(C_0,C_1,\lambda_0,d,\nu)&=&C_0 \frac{d-\nu}{d-2\nu+1}J(C_0,C_1,\lambda_0,d,\nu+1)+\frac{\lambda_0}{d-2\nu+1}(C_0+C_1 \lambda_0^2)^{\frac{d}{2}-\nu}.
\eea
Integrals with factors of $\lambda$ in the numerator are obtained by using
\bea
J(C_0,C_1,\lambda_0,d,\nu,n=1)&\equiv& \int_0^\infty (\lambda-\lambda_0)^{n=1} (C_0+C_1(\lambda-\lambda_0)^2)^{-\nu+\frac{d}{2}}d\lambda\nonumber\\
&=& -\frac{1}{2\, C_1\,(\frac{d}{2} + 1 - \nu)} (C_0+C_1\lambda_0^2)^{\frac{d}{2} + 1 - \nu},
\eea
and the recurrence relations
\beq
J(C_0,C_1,\lambda_0,d,\nu,n)=\frac{1}{C_1}(J(C_0,C_1,\lambda_0,d,\nu-1,n-1)-C_0 J(C_0,C_1,\lambda_0,d,\nu,n-2)).
\eeq
For   convenience in some of the calculations for the currents, the following integral is defined:
\beq
\tilde{J}(C_0,C_1,\lambda_0,d,\nu,1)\equiv J(C_0,C_1,\lambda_0,d,\nu,1)+\lambda_0 J(C_0,C_1,\lambda_0,d,\nu)
\eeq
 
 For the calculations in this work the following  integrals are needed at $d=4-2\eps$:
 \bea
J(C_0,C_1,\lambda_0,d,3)&=&\frac{1}{\sqrt{C_0 C_1}}\left(\frac{\pi}{2}+\arctan( \lambda_0\sqrt{\frac{C_1}{C_0}})\right) \nonumber\\
J(C_0,C_1,\lambda_0,d,2)&=&\frac{1}{d-3}(\lambda_0(C_0+C_1\lambda_0^2)^{\frac{d}{2}-2}+(d-4)C_0 J(C_0,C_1,\lambda_0,d,3))\nonumber\\
J(C_0,C_1,\lambda_0,d,1)&=&\frac{1}{d-1}(\lambda_0(C_0+C_1\lambda_0^2)^{\frac{d}{2}-1}+(d-2)J(C_0,C_1,\lambda_0,d,2))
 \eea

\subsection*{Specific integrals}
Here a summary of relevant one-loop integrals for the calculations in this work is provided for the convenience of the reader.

1) Loop integrals involving only relativistic propagators
\bea
I(0,1,M)&=&-\frac{i}{(4\pi)^{\frac d2}}\Gamma(1-\frac d2) M^{d-2}\nonumber\\
I(0,2,M)&=&\frac{i}{(4\pi)^{\frac d2}}\Gamma(2-\frac d2) M^{d-4}\nonumber\\
K(q,M_a,M_b)&\equiv&\int \widetilde{d^dk} \frac{1}{(k^2-M_a^2+i\eps)((k+q)^2-M_b^2+i\eps)}=\int_0^1 d\alpha \; I(0,2,\Lambda(\alpha))\nonumber\\
K^\mu(q,M_a,M_b)&\equiv&\int \widetilde{d^dk} \frac{k^\mu}{(k^2-M_a^2+i\eps)((k+q)^2-M_b^2+i\eps)}=\int_0^1 d\alpha \; (\alpha-1)\, q^\mu\,I(0,2,\Lambda(\alpha))\nonumber\\
K^{\mu\nu}(q,M_a,M_b)&\equiv&\int \widetilde{d^dk} \frac{k^\mu k^\nu}{(k^2-M_a^2+i\eps)((k+q)^2-M_b^2+i\eps)}\nonumber\\
&=&\int_0^1 d\alpha \;( (1-\alpha)^2\, q^\mu q^\nu\,I(0,2,\Lambda(\alpha))+\frac{g^{\mu\nu}}{d} I(1,2,\Lambda(\alpha))),
\eea
where:
\beq
\Lambda(\alpha)=\sqrt{\alpha M_a^2+(1-\alpha)M_b^2-\alpha(1-\alpha) q^2}\nonumber
\eeq
2) Loop integrals involving one heavy propagator
\bea
H(p_0,M)&\equiv&\int \widetilde{d^dk} \frac{1}{(p_0-k_0+i\eps)(k^2-M^2+i\eps)}\nonumber\\
&=&\frac{2i}{(4\pi)^{\frac d2}}\Gamma(2-\frac d2) J(M^2-p_0^{   2},1,p_0,d,2)   \nonumber\\
H^{ij}(p_0,M)&\equiv& \int \widetilde{d^dk} \frac{k^i k^j}{(p_0-k_0+i\eps)(k^2-M^2+i\eps)}\nonumber\\
&=& -\frac{i}{(4\pi)^{\frac d2}} g^{ij}\Gamma(1-\frac d2)J(M^2-p_0^2,1,p_0,d,1)
 \\
  H^{ij\mu}(p_0,q,M_a,M_b)&\equiv&\int \widetilde{d^dk} \frac{k^i (k+q)^j (2k+q)^\mu}{(p_0-k_0+i\eps)(k^2-M_a^2+i\eps)((k+q)^2-M_b^2+i\eps)}\nonumber\\
  &=&i\frac{4}{(4\pi)^{\frac d2}}\int_0^1 d\alpha\left\{ -\frac 12 \Gamma(3-\frac d2) q^iq^j \alpha(1-\alpha)\right.\nonumber\\
  &\times& \left((1-2\alpha)q^\mu J(C_0,C_1,\lambda_0,d,3)-2\,g^{\mu 0}  \tilde{J}(C_0,C_1,\lambda_0,d,3,1)\right)\nonumber\\
  &+& \Gamma(2-\frac d2)\left((-(1-2\alpha)g^{ij} q^\mu+2(\alpha g^{\mu i} q^j-(1-\alpha) g^{\mu j} q^i)) J(C_0,C_1,\lambda_0,d,2)\right.\nonumber\\
&+&  \left.\left. 2 g^{ij} g^{\mu 0} \tilde{J}(C_0,C_1,\lambda_0,d,2,1)\right)\right\},\nonumber
\eea
where:
\bea
C_0&=& \alpha M_a^2+(1-\alpha) M_b^2-p_0^2-2(1-\alpha) p_0 q_0-(1-\alpha)(\alpha\, q^2+(1-\alpha)q_0^2)\nonumber\\
C_1&=& 1\nonumber\\
\lambda_0&=& p_0+(1-\alpha) q_0.
\eea
The polynomial pieces of the integrals are as follows:
\bea
H(p_0,M)^{\text{poly}}&=&\frac{i}{(4\pi)^2} 2 p_0(\lambda_\eps+2)\nonumber\\
H^{ij}(p_0,M)^{\text{poly}}&=&\frac{i}{(4\pi)^2} \frac{p_0}{3}((3M^2-2p_0^2)\lambda_\eps+7 M^2-\frac{16}{3}p_0^2)\nonumber\\
H^{ij\mu}(p_0,q,M_a,M_b)^{\text{poly}}&=&\frac{i}{96 \pi ^2}
 \left(
 \lambda _\epsilon  
 \left(
 g^{ij} 
 \left(
 g^{\mu 0} 
 \left(
 -3 (M_a^2+
   M_b^2)+12 \, p_0(p_0+ q_0)+q^2+4 q_0^2
\right)
-q_0 q^\mu 
 \right)\right.\right.
 \nonumber\\
   &-&
   \left.
   2 q^i (3 p_0+2 q_0) g^{\mu j}+2 q^j ((3p_0+q_0) g^{\mu i}+q^i g^{\mu 0})
\right)\nonumber\\
   &+&  g^{ij}
   \left(
   g^{\mu 0} 
   \left(
   -3
   (M_a^2+ M_b^2)+24\, p_0( p_0+ q_0)+q^2+8
   q_0^2
   \right)
   -2 q_0 q^\mu 
   \right)
   \nonumber\\
   &-&
   \left. 
   4 q^i (3 p_0+2 q_0) g^{\mu j}+4 q^j (3 p_0+q_0) g^{\mu i}
   \right),
\eea
 where the UV divergence is given by the terms proportional to $\lambda_\eps\equiv 1/\eps-\gamma+\log 4\pi$, where $d=4-2\eps$.

\section{Interaction  and vector current vertices needed in loop calculations}
\label{app:vertices}

The interaction   and   currents vertices needed in the one-loop calculations are given for completeness.
~\vspace*{-.5cm}

\begin{center}
\begin{figure}[h!!]
\centerline{\includegraphics[width=12.cm,angle=-0]{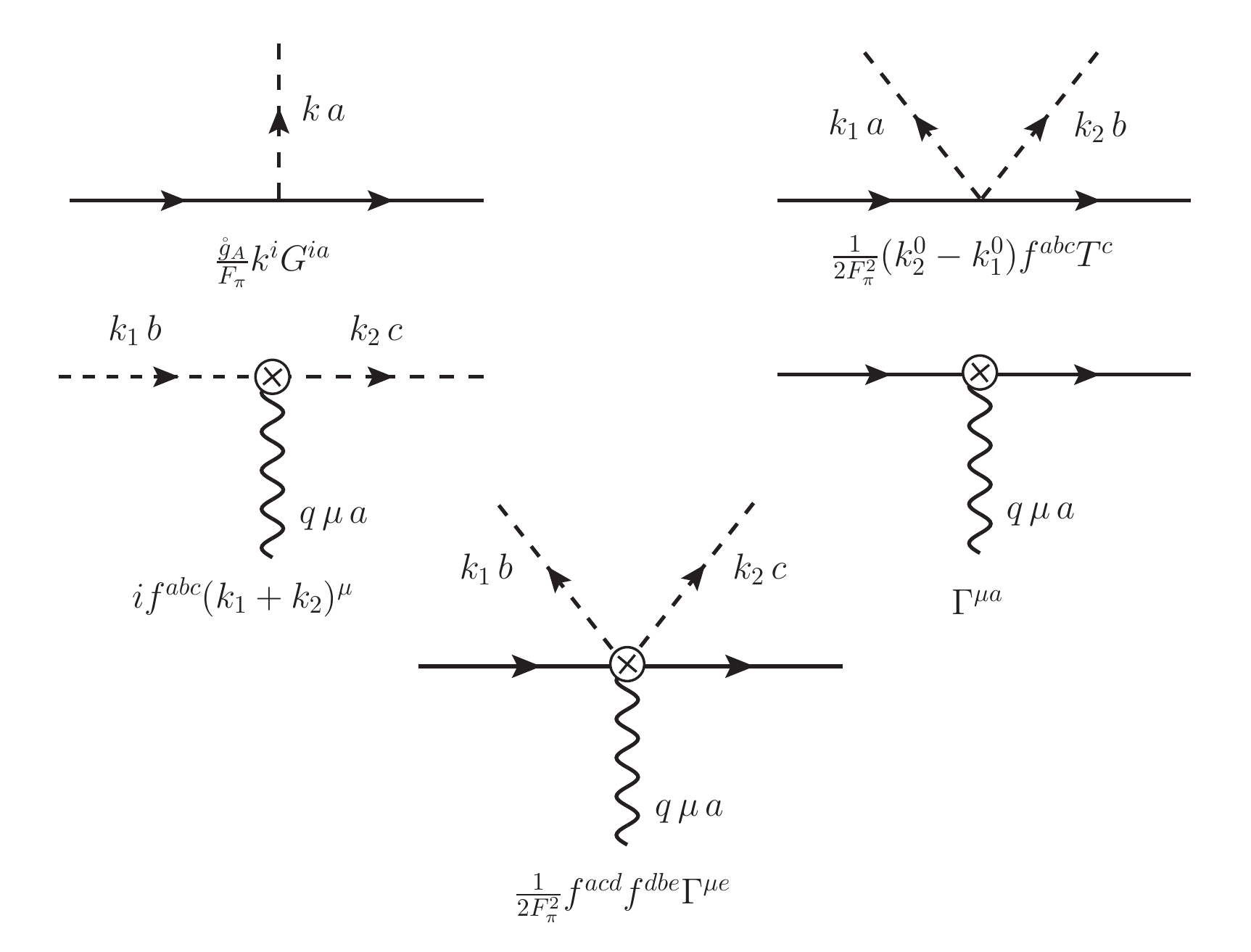}}
\caption{ The vector current vertices indicated with a square are the magnetic ones. The momentum $q$ is incoming, and $\Gamma^{\mu a}=g^{\mu 0} T^a+i\frac{\kappa}{\Lambda} \,\eps^{0\mu i j} \,f^{abc}f^{cbd}\,q^i \,G^{jd}$. }
\label{fig:vertices}
\end{figure}
\end{center}

\newpage
 
\bibliography{Refs}

\newcommand{\noop}[1]{}
\begin{thebibliography}{45}
\expandafter\ifx\csname natexlab\endcsname\relax\def\natexlab#1{#1}\fi
\expandafter\ifx\csname bibnamefont\endcsname\relax
  \def\bibnamefont#1{#1}\fi
\expandafter\ifx\csname bibfnamefont\endcsname\relax
  \def\bibfnamefont#1{#1}\fi
\expandafter\ifx\csname citenamefont\endcsname\relax
  \def\citenamefont#1{#1}\fi
\expandafter\ifx\csname url\endcsname\relax
  \def\url#1{\texttt{#1}}\fi
\expandafter\ifx\csname urlprefix\endcsname\relax\def\urlprefix{URL }\fi
\providecommand{\bibinfo}[2]{#2}
\providecommand{\eprint}[2][]{\url{#2}}

\bibitem[{\citenamefont{Ye et~al.}(2018)\citenamefont{Ye, Arrington, Hill, and
  Lee}}]{Ye:2017gyb}
\bibinfo{author}{\bibfnamefont{Z.}~\bibnamefont{Ye}},
  \bibinfo{author}{\bibfnamefont{J.}~\bibnamefont{Arrington}},
  \bibinfo{author}{\bibfnamefont{R.~J.} \bibnamefont{Hill}}, \bibnamefont{and}
  \bibinfo{author}{\bibfnamefont{G.}~\bibnamefont{Lee}},
  \bibinfo{journal}{Phys. Lett.} \textbf{\bibinfo{volume}{B777}},
  \bibinfo{pages}{8} (\bibinfo{year}{2018}), \eprint{1707.09063}.

\bibitem[{\citenamefont{Cabibbo et~al.}(2003)\citenamefont{Cabibbo, Swallow,
  and Winston}}]{Cabibbo:2003cu}
\bibinfo{author}{\bibfnamefont{N.}~\bibnamefont{Cabibbo}},
  \bibinfo{author}{\bibfnamefont{E.~C.} \bibnamefont{Swallow}},
  \bibnamefont{and} \bibinfo{author}{\bibfnamefont{R.}~\bibnamefont{Winston}},
  \bibinfo{journal}{Ann. Rev. Nucl. Part. Sci.} \textbf{\bibinfo{volume}{53}},
  \bibinfo{pages}{39} (\bibinfo{year}{2003}), \eprint{hep-ph/0307298}.

\bibitem[{\citenamefont{Aubin et~al.}(2009)\citenamefont{Aubin, Orginos,
  Pascalutsa, and Vanderhaeghen}}]{Aubin:2008qp}
\bibinfo{author}{\bibfnamefont{C.}~\bibnamefont{Aubin}},
  \bibinfo{author}{\bibfnamefont{K.}~\bibnamefont{Orginos}},
  \bibinfo{author}{\bibfnamefont{V.}~\bibnamefont{Pascalutsa}},
  \bibnamefont{and}
  \bibinfo{author}{\bibfnamefont{M.}~\bibnamefont{Vanderhaeghen}},
  \bibinfo{journal}{Phys. Rev.} \textbf{\bibinfo{volume}{D79}},
  \bibinfo{pages}{051502(R)} (\bibinfo{year}{2009}), \eprint{0811.2440}.

\bibitem[{\citenamefont{Shanahan et~al.}(2015)\citenamefont{Shanahan, Cooke,
  Horsley, Nakamura, Rakow, Schierholz, Thomas, Young, and
  Zanotti}}]{Shanahan:2015dka}
\bibinfo{author}{\bibfnamefont{P.~E.} \bibnamefont{Shanahan}},
  \bibinfo{author}{\bibfnamefont{A.~N.} \bibnamefont{Cooke}},
  \bibinfo{author}{\bibfnamefont{R.}~\bibnamefont{Horsley}},
  \bibinfo{author}{\bibfnamefont{Y.}~\bibnamefont{Nakamura}},
  \bibinfo{author}{\bibfnamefont{P.~E.~L.} \bibnamefont{Rakow}},
  \bibinfo{author}{\bibfnamefont{G.}~\bibnamefont{Schierholz}},
  \bibinfo{author}{\bibfnamefont{A.~W.} \bibnamefont{Thomas}},
  \bibinfo{author}{\bibfnamefont{R.~D.} \bibnamefont{Young}}, \bibnamefont{and}
  \bibinfo{author}{\bibfnamefont{J.~M.} \bibnamefont{Zanotti}},
  \bibinfo{journal}{Phys. Rev.} \textbf{\bibinfo{volume}{D92}},
  \bibinfo{pages}{074029} (\bibinfo{year}{2015}), \eprint{1508.06923}.

\bibitem[{\citenamefont{Parre\~no et~al.}(2017)\citenamefont{Parre\~no, Savage,
  Tiburzi, Wilhelm, Chang, Detmold, and Orginos}}]{Parreno:2016fwu}
\bibinfo{author}{\bibfnamefont{A.}~\bibnamefont{Parre\~no}},
  \bibinfo{author}{\bibfnamefont{M.~J.} \bibnamefont{Savage}},
  \bibinfo{author}{\bibfnamefont{B.~C.} \bibnamefont{Tiburzi}},
  \bibinfo{author}{\bibfnamefont{J.}~\bibnamefont{Wilhelm}},
  \bibinfo{author}{\bibfnamefont{E.}~\bibnamefont{Chang}},
  \bibinfo{author}{\bibfnamefont{W.}~\bibnamefont{Detmold}}, \bibnamefont{and}
  \bibinfo{author}{\bibfnamefont{K.}~\bibnamefont{Orginos}},
  \bibinfo{journal}{Phys. Rev.} \textbf{\bibinfo{volume}{D95}},
  \bibinfo{pages}{114513} (\bibinfo{year}{2017}), \eprint{1609.03985}.

\bibitem[{\citenamefont{Alexandrou et~al.}(2018)\citenamefont{Alexandrou,
  Constantinou, Hadjiyiannakou, Jansen, Kallidonis, Koutsou, and Vaquero
  Avilés-Casco}}]{Alexandrou:2018zdf}
\bibinfo{author}{\bibfnamefont{C.}~\bibnamefont{Alexandrou}},
  \bibinfo{author}{\bibfnamefont{M.}~\bibnamefont{Constantinou}},
  \bibinfo{author}{\bibfnamefont{K.}~\bibnamefont{Hadjiyiannakou}},
  \bibinfo{author}{\bibfnamefont{K.}~\bibnamefont{Jansen}},
  \bibinfo{author}{\bibfnamefont{C.}~\bibnamefont{Kallidonis}},
  \bibinfo{author}{\bibfnamefont{G.}~\bibnamefont{Koutsou}}, \bibnamefont{and}
  \bibinfo{author}{\bibfnamefont{A.}~\bibnamefont{Vaquero Avilés-Casco}},
  \bibinfo{journal}{Phys. Rev.} \textbf{\bibinfo{volume}{D97}},
  \bibinfo{pages}{094504} (\bibinfo{year}{2018}), \eprint{1801.09581}.

\bibitem[{\citenamefont{Jenkins et~al.}(1993)\citenamefont{Jenkins, Luke,
  Manohar, and Savage}}]{Jenkins:1992pi}
\bibinfo{author}{\bibfnamefont{E.~E.} \bibnamefont{Jenkins}},
  \bibinfo{author}{\bibfnamefont{M.~E.} \bibnamefont{Luke}},
  \bibinfo{author}{\bibfnamefont{A.~V.} \bibnamefont{Manohar}},
  \bibnamefont{and} \bibinfo{author}{\bibfnamefont{M.~J.}
  \bibnamefont{Savage}}, \bibinfo{journal}{Phys. Lett.}
  \textbf{\bibinfo{volume}{B302}}, \bibinfo{pages}{482} (\bibinfo{year}{1993}),
  \bibinfo{note}{[Erratum: Phys. Lett.B388,866(1996)]},
  \eprint{hep-ph/9212226}.

\bibitem[{\citenamefont{Butler et~al.}(1994)\citenamefont{Butler, Savage, and
  Springer}}]{Butler_1994}
\bibinfo{author}{\bibfnamefont{M.~N.} \bibnamefont{Butler}},
  \bibinfo{author}{\bibfnamefont{M.~J.} \bibnamefont{Savage}},
  \bibnamefont{and} \bibinfo{author}{\bibfnamefont{R.~P.}
  \bibnamefont{Springer}}, \bibinfo{journal}{Physical Review D}
  \textbf{\bibinfo{volume}{49}}, \bibinfo{pages}{3459–3465}
  (\bibinfo{year}{1994}), ISSN \bibinfo{issn}{0556-2821},
  \urlprefix\url{http://dx.doi.org/10.1103/PhysRevD.49.3459}.

\bibitem[{\citenamefont{Durand and Ha}(1998)}]{Durand:1997ya}
\bibinfo{author}{\bibfnamefont{L.}~\bibnamefont{Durand}} \bibnamefont{and}
  \bibinfo{author}{\bibfnamefont{P.}~\bibnamefont{Ha}}, \bibinfo{journal}{Phys.
  Rev.} \textbf{\bibinfo{volume}{D58}}, \bibinfo{pages}{013010}
  (\bibinfo{year}{1998}), \eprint{hep-ph/9712492}.

\bibitem[{\citenamefont{Bernard et~al.}(1998)\citenamefont{Bernard, Fearing,
  Hemmert, and Meissner}}]{Bernard:1998gv}
\bibinfo{author}{\bibfnamefont{V.}~\bibnamefont{Bernard}},
  \bibinfo{author}{\bibfnamefont{H.~W.} \bibnamefont{Fearing}},
  \bibinfo{author}{\bibfnamefont{T.~R.} \bibnamefont{Hemmert}},
  \bibnamefont{and} \bibinfo{author}{\bibfnamefont{U.-G.}
  \bibnamefont{Meissner}}, \bibinfo{journal}{Nucl. Phys.}
  \textbf{\bibinfo{volume}{A635}}, \bibinfo{pages}{121} (\bibinfo{year}{1998}),
  \bibinfo{note}{[Erratum: Nucl. Phys.A642,563(1998)]},
  \eprint{hep-ph/9801297}.

\bibitem[{\citenamefont{Pascalutsa and
  Vanderhaeghen}(2005)}]{Pascalutsa:2004je}
\bibinfo{author}{\bibfnamefont{V.}~\bibnamefont{Pascalutsa}} \bibnamefont{and}
  \bibinfo{author}{\bibfnamefont{M.}~\bibnamefont{Vanderhaeghen}},
  \bibinfo{journal}{Phys. Rev. Lett.} \textbf{\bibinfo{volume}{94}},
  \bibinfo{pages}{102003} (\bibinfo{year}{2005}), \eprint{nucl-th/0412113}.

\bibitem[{\citenamefont{Ledwig et~al.}(2012)\citenamefont{Ledwig,
  Martin-Camalich, Pascalutsa, and Vanderhaeghen}}]{Ledwig:2011cx}
\bibinfo{author}{\bibfnamefont{T.}~\bibnamefont{Ledwig}},
  \bibinfo{author}{\bibfnamefont{J.}~\bibnamefont{Martin-Camalich}},
  \bibinfo{author}{\bibfnamefont{V.}~\bibnamefont{Pascalutsa}},
  \bibnamefont{and}
  \bibinfo{author}{\bibfnamefont{M.}~\bibnamefont{Vanderhaeghen}},
  \bibinfo{journal}{Phys. Rev.} \textbf{\bibinfo{volume}{D85}},
  \bibinfo{pages}{034013} (\bibinfo{year}{2012}), \eprint{1108.2523}.

\bibitem[{\citenamefont{Geng et~al.}(2009{\natexlab{a}})\citenamefont{Geng,
  Martin~Camalich, and Vicente~Vacas}}]{Geng:2009ys}
\bibinfo{author}{\bibfnamefont{L.~S.} \bibnamefont{Geng}},
  \bibinfo{author}{\bibfnamefont{J.}~\bibnamefont{Martin~Camalich}},
  \bibnamefont{and} \bibinfo{author}{\bibfnamefont{M.~J.}
  \bibnamefont{Vicente~Vacas}}, \bibinfo{journal}{Phys. Rev.}
  \textbf{\bibinfo{volume}{D80}}, \bibinfo{pages}{034027}
  (\bibinfo{year}{2009}{\natexlab{a}}), \eprint{0907.0631}.

\bibitem[{\citenamefont{Geng et~al.}(2009{\natexlab{b}})\citenamefont{Geng,
  Martin~Camalich, and Vicente~Vacas}}]{Geng:2009hh}
\bibinfo{author}{\bibfnamefont{L.~S.} \bibnamefont{Geng}},
  \bibinfo{author}{\bibfnamefont{J.}~\bibnamefont{Martin~Camalich}},
  \bibnamefont{and} \bibinfo{author}{\bibfnamefont{M.~J.}
  \bibnamefont{Vicente~Vacas}}, \bibinfo{journal}{Phys. Lett.}
  \textbf{\bibinfo{volume}{B676}}, \bibinfo{pages}{63}
  (\bibinfo{year}{2009}{\natexlab{b}}), \eprint{0903.0779}.

\bibitem[{\citenamefont{Jiang and Tiburzi}(2010)}]{Jiang:2009jn}
\bibinfo{author}{\bibfnamefont{F.-J.} \bibnamefont{Jiang}} \bibnamefont{and}
  \bibinfo{author}{\bibfnamefont{B.~C.} \bibnamefont{Tiburzi}},
  \bibinfo{journal}{Phys. Rev.} \textbf{\bibinfo{volume}{D81}},
  \bibinfo{pages}{034017} (\bibinfo{year}{2010}), \eprint{0912.2077}.

\bibitem[{\citenamefont{Luty et~al.}(1995)\citenamefont{Luty, March-Russell,
  and White}}]{Luty:1994ub}
\bibinfo{author}{\bibfnamefont{M.~A.} \bibnamefont{Luty}},
  \bibinfo{author}{\bibfnamefont{J.}~\bibnamefont{March-Russell}},
  \bibnamefont{and} \bibinfo{author}{\bibfnamefont{M.~J.} \bibnamefont{White}},
  \bibinfo{journal}{Phys. Rev.} \textbf{\bibinfo{volume}{D51}},
  \bibinfo{pages}{2332} (\bibinfo{year}{1995}), \eprint{hep-ph/9405272}.

\bibitem[{\citenamefont{Flores-Mendieta}(2009)}]{FloresMendieta:2009rq}
\bibinfo{author}{\bibfnamefont{R.}~\bibnamefont{Flores-Mendieta}},
  \bibinfo{journal}{Phys. Rev.} \textbf{\bibinfo{volume}{D80}},
  \bibinfo{pages}{094014} (\bibinfo{year}{2009}), \eprint{0910.1103}.

\bibitem[{\citenamefont{Ahuatzin et~al.}(2014)\citenamefont{Ahuatzin,
  Flores-Mendieta, Hernandez-Ruiz, and Hofmann}}]{Ahuatzin:2010ef}
\bibinfo{author}{\bibfnamefont{G.}~\bibnamefont{Ahuatzin}},
  \bibinfo{author}{\bibfnamefont{R.}~\bibnamefont{Flores-Mendieta}},
  \bibinfo{author}{\bibfnamefont{M.~A.} \bibnamefont{Hernandez-Ruiz}},
  \bibnamefont{and} \bibinfo{author}{\bibfnamefont{C.~P.}
  \bibnamefont{Hofmann}}, \bibinfo{journal}{Phys. Rev.}
  \textbf{\bibinfo{volume}{D89}}, \bibinfo{pages}{034012}
  (\bibinfo{year}{2014}), \eprint{1011.5268}.

\bibitem[{\citenamefont{Jenkins}(2012)}]{Jenkins:2011dr}
\bibinfo{author}{\bibfnamefont{E.~E.} \bibnamefont{Jenkins}},
  \bibinfo{journal}{Phys. Rev.} \textbf{\bibinfo{volume}{D85}},
  \bibinfo{pages}{065007} (\bibinfo{year}{2012}), \eprint{1111.2055}.

\bibitem[{\citenamefont{Flores-Mendieta and
  Goity}(2014)}]{Flores-Mendieta:2014vaa}
\bibinfo{author}{\bibfnamefont{R.}~\bibnamefont{Flores-Mendieta}}
  \bibnamefont{and} \bibinfo{author}{\bibfnamefont{J.~L.} \bibnamefont{Goity}},
  \bibinfo{journal}{Phys. Rev.} \textbf{\bibinfo{volume}{D90}},
  \bibinfo{pages}{114008} (\bibinfo{year}{2014}), \eprint{1407.0926}.

\bibitem[{\citenamefont{Flores-Mendieta and
  Rivera-Ruiz}(2015)}]{Flores-Mendieta:2015wir}
\bibinfo{author}{\bibfnamefont{R.}~\bibnamefont{Flores-Mendieta}}
  \bibnamefont{and} \bibinfo{author}{\bibfnamefont{M.~A.}
  \bibnamefont{Rivera-Ruiz}}, \bibinfo{journal}{Phys. Rev.}
  \textbf{\bibinfo{volume}{D92}}, \bibinfo{pages}{094026}
  (\bibinfo{year}{2015}), \eprint{1511.02932}.

\bibitem[{\citenamefont{Jenkins}(1996)}]{Jenkins:1995gc}
\bibinfo{author}{\bibfnamefont{E.~E.} \bibnamefont{Jenkins}},
  \bibinfo{journal}{Phys. Rev.} \textbf{\bibinfo{volume}{D53}},
  \bibinfo{pages}{2625} (\bibinfo{year}{1996}), \eprint{hep-ph/9509433}.

\bibitem[{\citenamefont{Calle~Cordon and Goity}(2013)}]{CalleCordon:2012xz}
\bibinfo{author}{\bibfnamefont{A.}~\bibnamefont{Calle~Cordon}}
  \bibnamefont{and} \bibinfo{author}{\bibfnamefont{J.~L.} \bibnamefont{Goity}},
  \bibinfo{journal}{Phys. Rev.} \textbf{\bibinfo{volume}{D87}},
  \bibinfo{pages}{016019} (\bibinfo{year}{2013}), \eprint{1210.2364}.

\bibitem[{\citenamefont{Fernando and Goity}(2018)}]{Fernando:2017yqd}
\bibinfo{author}{\bibfnamefont{I.~P.} \bibnamefont{Fernando}} \bibnamefont{and}
  \bibinfo{author}{\bibfnamefont{J.~L.} \bibnamefont{Goity}},
  \bibinfo{journal}{Phys. Rev.} \textbf{\bibinfo{volume}{D97}},
  \bibinfo{pages}{054010} (\bibinfo{year}{2018}), \eprint{1712.01672}.

\bibitem[{\citenamefont{Fernando et~al.}(2018)\citenamefont{Fernando,
  Alarc\'on, and Goity}}]{Fernando:2018jrz}
\bibinfo{author}{\bibfnamefont{I.~P.} \bibnamefont{Fernando}},
  \bibinfo{author}{\bibfnamefont{J.~M.} \bibnamefont{Alarc\'on}},
  \bibnamefont{and} \bibinfo{author}{\bibfnamefont{J.~L.} \bibnamefont{Goity}},
  \bibinfo{journal}{Phys. Lett.} \textbf{\bibinfo{volume}{B781}},
  \bibinfo{pages}{719} (\bibinfo{year}{2018}), \eprint{1804.03094}.

\bibitem[{\citenamefont{Fernando and Goity}(2019)}]{Fernando:2019sfm}
\bibinfo{author}{\bibfnamefont{I.~P.} \bibnamefont{Fernando}} \bibnamefont{and}
  \bibinfo{author}{\bibfnamefont{J.~L.} \bibnamefont{Goity}}, in
  \emph{\bibinfo{booktitle}{{9th International Workshop on Chiral Dynamics
  (CD18) Durham, NC, USA, September 17-21, 2018}}} (\bibinfo{year}{2019}),
  \eprint{1904.07112},
  \urlprefix\url{https://misportal.jlab.org/ul/publications/view_pub.cfm?pub_id=15886}.

\bibitem[{\citenamefont{Jenkins and
  Manohar}(1994{\natexlab{a}})}]{Jenkins_1994}
\bibinfo{author}{\bibfnamefont{E.}~\bibnamefont{Jenkins}} \bibnamefont{and}
  \bibinfo{author}{\bibfnamefont{A.~V.} \bibnamefont{Manohar}},
  \bibinfo{journal}{Physics Letters B} \textbf{\bibinfo{volume}{335}},
  \bibinfo{pages}{452–459} (\bibinfo{year}{1994}{\natexlab{a}}), ISSN
  \bibinfo{issn}{0370-2693},
  \urlprefix\url{http://dx.doi.org/10.1016/0370-2693(94)90377-8}.

\bibitem[{\citenamefont{Dai et~al.}(1996)\citenamefont{Dai, Dashen, Jenkins,
  and Manohar}}]{Dai:1995zg}
\bibinfo{author}{\bibfnamefont{J.}~\bibnamefont{Dai}},
  \bibinfo{author}{\bibfnamefont{R.~F.} \bibnamefont{Dashen}},
  \bibinfo{author}{\bibfnamefont{E.~E.} \bibnamefont{Jenkins}},
  \bibnamefont{and} \bibinfo{author}{\bibfnamefont{A.~V.}
  \bibnamefont{Manohar}}, \bibinfo{journal}{Phys. Rev.}
  \textbf{\bibinfo{volume}{D53}}, \bibinfo{pages}{273} (\bibinfo{year}{1996}),
  \eprint{hep-ph/9506273}.

\bibitem[{\citenamefont{Buchmann and Lebed}(2003)}]{Buchmann:2002et}
\bibinfo{author}{\bibfnamefont{A.~J.} \bibnamefont{Buchmann}} \bibnamefont{and}
  \bibinfo{author}{\bibfnamefont{R.~F.} \bibnamefont{Lebed}},
  \bibinfo{journal}{Phys. Rev.} \textbf{\bibinfo{volume}{D67}},
  \bibinfo{pages}{016002} (\bibinfo{year}{2003}), \eprint{hep-ph/0207358}.

\bibitem[{\citenamefont{Lebed and Martin}(2004)}]{Lebed:2004fj}
\bibinfo{author}{\bibfnamefont{R.~F.} \bibnamefont{Lebed}} \bibnamefont{and}
  \bibinfo{author}{\bibfnamefont{D.~R.} \bibnamefont{Martin}},
  \bibinfo{journal}{Phys. Rev.} \textbf{\bibinfo{volume}{D70}},
  \bibinfo{pages}{016008} (\bibinfo{year}{2004}), \eprint{hep-ph/0404160}.

\bibitem[{\citenamefont{Hammer and Meissner}(2004)}]{Hammer:2003ai}
\bibinfo{author}{\bibfnamefont{H.~W.} \bibnamefont{Hammer}} \bibnamefont{and}
  \bibinfo{author}{\bibfnamefont{U.-G.} \bibnamefont{Meissner}},
  \bibinfo{journal}{Eur. Phys. J.} \textbf{\bibinfo{volume}{A20}},
  \bibinfo{pages}{469} (\bibinfo{year}{2004}), \eprint{hep-ph/0312081}.

\bibitem[{\citenamefont{Belushkin et~al.}(2006)\citenamefont{Belushkin, Hammer,
  and Meissner}}]{Belushkin:2005ds}
\bibinfo{author}{\bibfnamefont{M.~A.} \bibnamefont{Belushkin}},
  \bibinfo{author}{\bibfnamefont{H.~W.} \bibnamefont{Hammer}},
  \bibnamefont{and} \bibinfo{author}{\bibfnamefont{U.-G.}
  \bibnamefont{Meissner}}, \bibinfo{journal}{Phys. Lett.}
  \textbf{\bibinfo{volume}{B633}}, \bibinfo{pages}{507} (\bibinfo{year}{2006}),
  \eprint{hep-ph/0510382}.

\bibitem[{\citenamefont{Belushkin et~al.}(2007)\citenamefont{Belushkin, Hammer,
  and Meissner}}]{Belushkin:2006qa}
\bibinfo{author}{\bibfnamefont{M.~A.} \bibnamefont{Belushkin}},
  \bibinfo{author}{\bibfnamefont{H.~W.} \bibnamefont{Hammer}},
  \bibnamefont{and} \bibinfo{author}{\bibfnamefont{U.-G.}
  \bibnamefont{Meissner}}, \bibinfo{journal}{Phys. Rev.}
  \textbf{\bibinfo{volume}{C75}}, \bibinfo{pages}{035202}
  (\bibinfo{year}{2007}), \eprint{hep-ph/0608337}.

\bibitem[{\citenamefont{Granados and Weiss}(2014)}]{Granados:2013moa}
\bibinfo{author}{\bibfnamefont{C.}~\bibnamefont{Granados}} \bibnamefont{and}
  \bibinfo{author}{\bibfnamefont{C.}~\bibnamefont{Weiss}},
  \bibinfo{journal}{JHEP} \textbf{\bibinfo{volume}{01}}, \bibinfo{pages}{092}
  (\bibinfo{year}{2014}), \eprint{1308.1634}.

\bibitem[{\citenamefont{Granados and Weiss}(2016)}]{Granados:2016jjl}
\bibinfo{author}{\bibfnamefont{C.}~\bibnamefont{Granados}} \bibnamefont{and}
  \bibinfo{author}{\bibfnamefont{C.}~\bibnamefont{Weiss}},
  \bibinfo{journal}{JHEP} \textbf{\bibinfo{volume}{06}}, \bibinfo{pages}{075}
  (\bibinfo{year}{2016}), \eprint{1603.08881}.

\bibitem[{\citenamefont{Alarc\'on et~al.}(2017)\citenamefont{Alarc\'on,
  Hiller~Blin, Vicente~Vacas, and Weiss}}]{Alarcon:2017asr}
\bibinfo{author}{\bibfnamefont{J.~M.} \bibnamefont{Alarc\'on}},
  \bibinfo{author}{\bibfnamefont{A.~N.} \bibnamefont{Hiller~Blin}},
  \bibinfo{author}{\bibfnamefont{M.~J.} \bibnamefont{Vicente~Vacas}},
  \bibnamefont{and} \bibinfo{author}{\bibfnamefont{C.}~\bibnamefont{Weiss}},
  \bibinfo{journal}{Nucl. Phys.} \textbf{\bibinfo{volume}{A964}},
  \bibinfo{pages}{18} (\bibinfo{year}{2017}), \eprint{1703.04534}.

\bibitem[{\citenamefont{Alarc\'on and
  Weiss}(2018{\natexlab{a}})}]{Alarcon:2017lhg}
\bibinfo{author}{\bibfnamefont{J.~M.} \bibnamefont{Alarc\'on}}
  \bibnamefont{and} \bibinfo{author}{\bibfnamefont{C.}~\bibnamefont{Weiss}},
  \bibinfo{journal}{Phys. Rev.} \textbf{\bibinfo{volume}{C97}},
  \bibinfo{pages}{055203} (\bibinfo{year}{2018}{\natexlab{a}}),
  \eprint{1710.06430}.

\bibitem[{\citenamefont{Alarc\'on and
  Weiss}(2018{\natexlab{b}})}]{Alarcon:2018irp}
\bibinfo{author}{\bibfnamefont{J.~M.} \bibnamefont{Alarc\'on}}
  \bibnamefont{and} \bibinfo{author}{\bibfnamefont{C.}~\bibnamefont{Weiss}},
  \bibinfo{journal}{Phys. Lett.} \textbf{\bibinfo{volume}{B784}},
  \bibinfo{pages}{373} (\bibinfo{year}{2018}{\natexlab{b}}),
  \eprint{1803.09748}.

\bibitem[{\citenamefont{Shrock}(1996)}]{Shrock:1995bp}
\bibinfo{author}{\bibfnamefont{R.}~\bibnamefont{Shrock}},
  \bibinfo{journal}{Phys. Rev.} \textbf{\bibinfo{volume}{D53}},
  \bibinfo{pages}{6465} (\bibinfo{year}{1996}), \eprint{hep-ph/9512430}.

\bibitem[{\citenamefont{Tanabashi et~al.}(2018)\citenamefont{Tanabashi,
  Hagiwara, Hikasa, Nakamura, Sumino, Takahashi, Tanaka, Agashe, Aielli, Amsler
  et~al.}}]{PhysRevD.98.030001}
\bibinfo{author}{\bibfnamefont{M.}~\bibnamefont{Tanabashi}},
  \bibinfo{author}{\bibfnamefont{K.}~\bibnamefont{Hagiwara}},
  \bibinfo{author}{\bibfnamefont{K.}~\bibnamefont{Hikasa}},
  \bibinfo{author}{\bibfnamefont{K.}~\bibnamefont{Nakamura}},
  \bibinfo{author}{\bibfnamefont{Y.}~\bibnamefont{Sumino}},
  \bibinfo{author}{\bibfnamefont{F.}~\bibnamefont{Takahashi}},
  \bibinfo{author}{\bibfnamefont{J.}~\bibnamefont{Tanaka}},
  \bibinfo{author}{\bibfnamefont{K.}~\bibnamefont{Agashe}},
  \bibinfo{author}{\bibfnamefont{G.}~\bibnamefont{Aielli}},
  \bibinfo{author}{\bibfnamefont{C.}~\bibnamefont{Amsler}},
  \bibnamefont{et~al.} (\bibinfo{collaboration}{Particle Data Group}),
  \bibinfo{journal}{Phys. Rev. D} \textbf{\bibinfo{volume}{98}},
  \bibinfo{pages}{030001} (\bibinfo{year}{2018}),
  \urlprefix\url{https://link.aps.org/doi/10.1103/PhysRevD.98.030001}.

\bibitem[{\citenamefont{Tiator et~al.}(2001)\citenamefont{Tiator, Drechsel,
  Hanstein, Kamalov, and Yang}}]{Tiator:2000iy}
\bibinfo{author}{\bibfnamefont{L.}~\bibnamefont{Tiator}},
  \bibinfo{author}{\bibfnamefont{D.}~\bibnamefont{Drechsel}},
  \bibinfo{author}{\bibfnamefont{O.}~\bibnamefont{Hanstein}},
  \bibinfo{author}{\bibfnamefont{S.~S.} \bibnamefont{Kamalov}},
  \bibnamefont{and} \bibinfo{author}{\bibfnamefont{S.~N.} \bibnamefont{Yang}},
  \bibinfo{journal}{Nucl. Phys.} \textbf{\bibinfo{volume}{A689}},
  \bibinfo{pages}{205} (\bibinfo{year}{2001}), \eprint{nucl-th/0012046}.

\bibitem[{\citenamefont{Dashen and Manohar}(1993)}]{Dashen:1993ac}
\bibinfo{author}{\bibfnamefont{R.~F.} \bibnamefont{Dashen}} \bibnamefont{and}
  \bibinfo{author}{\bibfnamefont{A.~V.} \bibnamefont{Manohar}},
  \bibinfo{journal}{Phys.Lett.} \textbf{\bibinfo{volume}{B315}},
  \bibinfo{pages}{438} (\bibinfo{year}{1993}), \eprint{hep-ph/9307242}.

\bibitem[{\citenamefont{Jenkins and
  Manohar}(1994{\natexlab{b}})}]{Jenkins:1994md}
\bibinfo{author}{\bibfnamefont{E.~E.} \bibnamefont{Jenkins}} \bibnamefont{and}
  \bibinfo{author}{\bibfnamefont{A.~V.} \bibnamefont{Manohar}},
  \bibinfo{journal}{Phys. Lett.} \textbf{\bibinfo{volume}{B335}},
  \bibinfo{pages}{452} (\bibinfo{year}{1994}{\natexlab{b}}),
  \eprint{hep-ph/9405431}.

\bibitem[{\citenamefont{Keller et~al.}(2011)}]{Keller:2011nt}
\bibinfo{author}{\bibfnamefont{D.}~\bibnamefont{Keller}} \bibnamefont{et~al.}
  (\bibinfo{collaboration}{CLAS}), \bibinfo{journal}{Phys. Rev.}
  \textbf{\bibinfo{volume}{D83}}, \bibinfo{pages}{072004}
  (\bibinfo{year}{2011}), \eprint{1103.5701}.

\bibitem[{\citenamefont{Keller et~al.}(2012)}]{Keller:2011aw}
\bibinfo{author}{\bibfnamefont{D.}~\bibnamefont{Keller}} \bibnamefont{et~al.}
  (\bibinfo{collaboration}{CLAS}), \bibinfo{journal}{Phys. Rev.}
  \textbf{\bibinfo{volume}{D85}}, \bibinfo{pages}{052004}
  (\bibinfo{year}{2012}), \eprint{1111.5444}.

\end{thebibliography}

\end{document}